\documentclass{IEEEoj}
\usepackage{graphicx}
\usepackage{svg}
\usepackage{comment}
\usepackage{bbding}
\usepackage{multirow}
\usepackage{longtable}
\usepackage{array}
\usepackage{url} 
\usepackage{cite}
\usepackage{float}
\usepackage{amsmath}
\usepackage{amssymb, amsfonts}
\usepackage{indentfirst}
\usepackage{verbatim}
\usepackage{makecell}
\usepackage{soul}
\usepackage{tabularx}
\usepackage{booktabs}
\usepackage{subcaption}
\usepackage{lettrine}
\usepackage{soul}
\usepackage{algorithm}
\usepackage{algorithmic}
\usepackage{colortbl} 
\usepackage{textcomp}
\usepackage[shortcuts,acronym, nonumberlist]{glossaries}
\newcolumntype{C}[1]{>{\centering\arraybackslash}m{#1}}

\def\BibTeX{{\rm B\kern-.05em{\sc i\kern-.025em b}\kern-.08em
    T\kern-.1667em\lower.7ex\hbox{E}\kern-.125emX}}
\AtBeginDocument{\definecolor{ojcolor}{cmyk}{0.93,0.59,0.15,0.02}}
\def\OJlogo{\vspace{-4pt}\hskip-4pt\includegraphics[height=18pt]{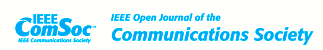}}

\makeglossaries 

\newacronym{uwb}{UWB}{Ultra-Wideband}
\newacronym{cir}{CIR}{Channel Impulse Response}
\newacronym{los}{LOS}{Line-of-Sight}
\newacronym{nlos}{NLOS}{Non-Line-of-Sight}
\newacronym{ssm}{SSM}{State Space Model}
\newacronym{mae}{MAE}{Mean Absolute Error}
\newacronym{mse}{MSE}{Mean Squared Error}
\newacronym{ssl}{SSL}{Self-Supervised Learning}
\newacronym{cnn}{CNN}{Convolutional Neural Network}
\newacronym{mae-model}{MAE}{Masked Autoencoder}
\newacronym{vit}{ViT}{Vision Transformer}
\newacronym{mlp}{MLP}{Multi-Layer Perceptron}
\newacronym{iq}{IQ}{In-phase and Quadrature}
\newacronym{csi}{CSI}{Channel State Information}
\newacronym{nlp}{NLP}{Natural Language Processing}
\newacronym{mimo}{MIMO}{Multiple-Input Multiple-Output}
\newacronym{ai}{AI}{Artificial Intelligence}
\newacronym{isac}{ISAC}{Integrated Sensing and Communication}
\newacronym{rf}{RF}{Radio Frequency}
\newacronym{peft}{PEFT}{Parameter-Efficient Fine-Tuning}
\newacronym{lora}{LoRA}{Low-Rank Adaptation}
\newacronym{ue}{UE}{User Equipment}
\newacronym{kd}{KD}{Knowledge Distillation}
\newacronym{wpfm}{WPFM}{Wireless Physical-Layer Foundation Models}
\newacronym{cfm}{CFM}{Channel Foundation Model}
\newacronym{twm}{TWM}{Telecom World Model}
\newacronym{nas}{NAS}{Neural Architecture Search}
\newacronym{llm}{LLM}{Large Language Model}
\newacronym{moe}{MoE}{Mixture of Experts}
\newacronym{awgn}{AWGN}{Additive White Gaussian Noise}
\newacronym{ofdm}{OFDM}{Orthogonal Frequency Division Multiplexing}
\newacronym{stft}{STFT}{Short-Time Fourier Transform}
\newacronym{fmcw}{FMCW}{Frequency-Modulated Continuous Wave}
\newacronym{bev}{BEV}{Bird's Eye View}
\newacronym{fdd}{FDD}{Frequency Division Duplex}
\newacronym{jepa}{JEPA}{Joint Embedding Predictive Architecture}
\newacronym{snn}{SNN}{Spiking Neural Network}
\newacronym{ttt}{TTT}{Test-Time Training}
\newacronym{rsrp}{RSRP}{Reference Signal Received Power}
\newacronym{har}{HAR}{Human Activity Recognition}
\newacronym{snr}{SNR}{Signal-to-Noise Ratio}
\newacronym{wwm}{WWM}{Wireless World Model}
\newacronym{eic}{EIC}{Environment-Intelligent Communication}
\newacronym{bs}{BS}{Base Station}
\newacronym{rope}{RoPE}{Rotary Positional Embedding}
\newacronym{wfm}{WFM}{Wireless Foundation Models}
\newacronym{rmse}{RMSE}{Root Mean Squared Error}
\newacronym{dnn}{DNN}{Deep Neural Network}
\newacronym{xr}{XR}{Extended Reality}

\makeatletter
\renewcommand{\p@subsection}{\thesection.}
\makeatother

\begin{document}

\receiveddate{XX Month, XXXX}
\reviseddate{XX Month, XXXX}
\accepteddate{XX Month, XXXX}
\publisheddate{XX Month, XXXX}
\currentdate{11 January, 2024}
\doiinfo{OJCOMS}

\title{Wireless Physical-Layer Foundation Models: Architectures, Learning Paradigms, Applications, and Deployment}

\author{Mohammad~Cheraghinia\IEEEauthorrefmark{1}, Davide~Buffelli\IEEEauthorrefmark{2}, Liu~Li\IEEEauthorrefmark{2}, 
Mohammad~Alhellani \IEEEauthorrefmark{1}
Jaron~Fontaine\IEEEauthorrefmark{1}, Sattar~Vakili\IEEEauthorrefmark{2},
Mamoun~Guenach\IEEEauthorrefmark{3},
Eli~De~Poorter\IEEEauthorrefmark{1} and~Adnan~Shahid\IEEEauthorrefmark{1}}
\affil{IDLab, Ghent University - imec, Ghent 9052, Belgium}
\affil{MediaTek Research, London, UK}
\affil{Interuniversity Microelectronics Centre (IMEC), 3000 Leuven, Belgium}
\corresp{CORRESPONDING AUTHOR: Mohammad~Cheraghinia (e-mail: mohammad.cheraghinia@ugent.be).}
\authornote{This work was supported by the Horizon Europe program under the MCSA Staff Exchanges 2021 grant agreement 101086218 (EVOLVE project); the Horizon-JU-SNS-2023 Research and Innovation Program under Grant Agreement No. 101139194 (6G-XCEL project); the Fund for Scientific Research Flanders (FWO-Vlaanderen) under SB-PhD Fellowship with grant number 1S52025N; and imec.icon project RAPIDNESS, which is co-financed by imec and Flanders Innovation and Entrepreneurship under project nr HBC.2024.0772.}
\markboth{Preparation of Papers for IEEE OPEN JOURNALS}{Cheraghinia \textit{et al.}}

\begin{abstract}
Foundation models, i.e., large neural networks pretrained on broad unlabeled data and adapted to many downstream tasks, have reshaped natural language processing and computer vision and are now being explored for the wireless physical layer. Wireless Physical-Layer Foundation Models (WPFMs) aim to learn transferable representations of signals such as channel state information (CSI), in-phase and quadrature (IQ) samples, and spectrograms so that a single pretrained backbone can support tasks ranging from channel estimation and prediction to localization and sensing while using limited task-specific data. This paper provides a dedicated review of WPFMs from learning design to practical deployment. We first establish the theoretical background, covering the neural architectures used for wireless signals, the self-supervised pretraining paradigms of masked modeling, contrastive learning, and generative pretraining, and the fine-tuning strategies that adapt pretrained models to downstream tasks. We then introduce a taxonomy that organizes existing models along five dimensions: architecture family, input modality and tokenization, pretraining objective, model scale and deployment target, and generalization capability. Building on this basis, we review applications across telecommunications, localization, and sensing, and, for each domain, analyze deployment feasibility by mapping model size to the memory, compute, and latency budgets of representative wireless hardware. Finally, we discuss model compression and efficient deployment, summarize the cross-cutting challenges, and outline open research directions. Our goal is to provide a reference that connects pretraining, architecture, and fine-tuning with the practical constraints of wireless systems.
\end{abstract}



\maketitle

\section{Introduction}

\begin{figure*}
    \centering
    \includegraphics[width=\linewidth]{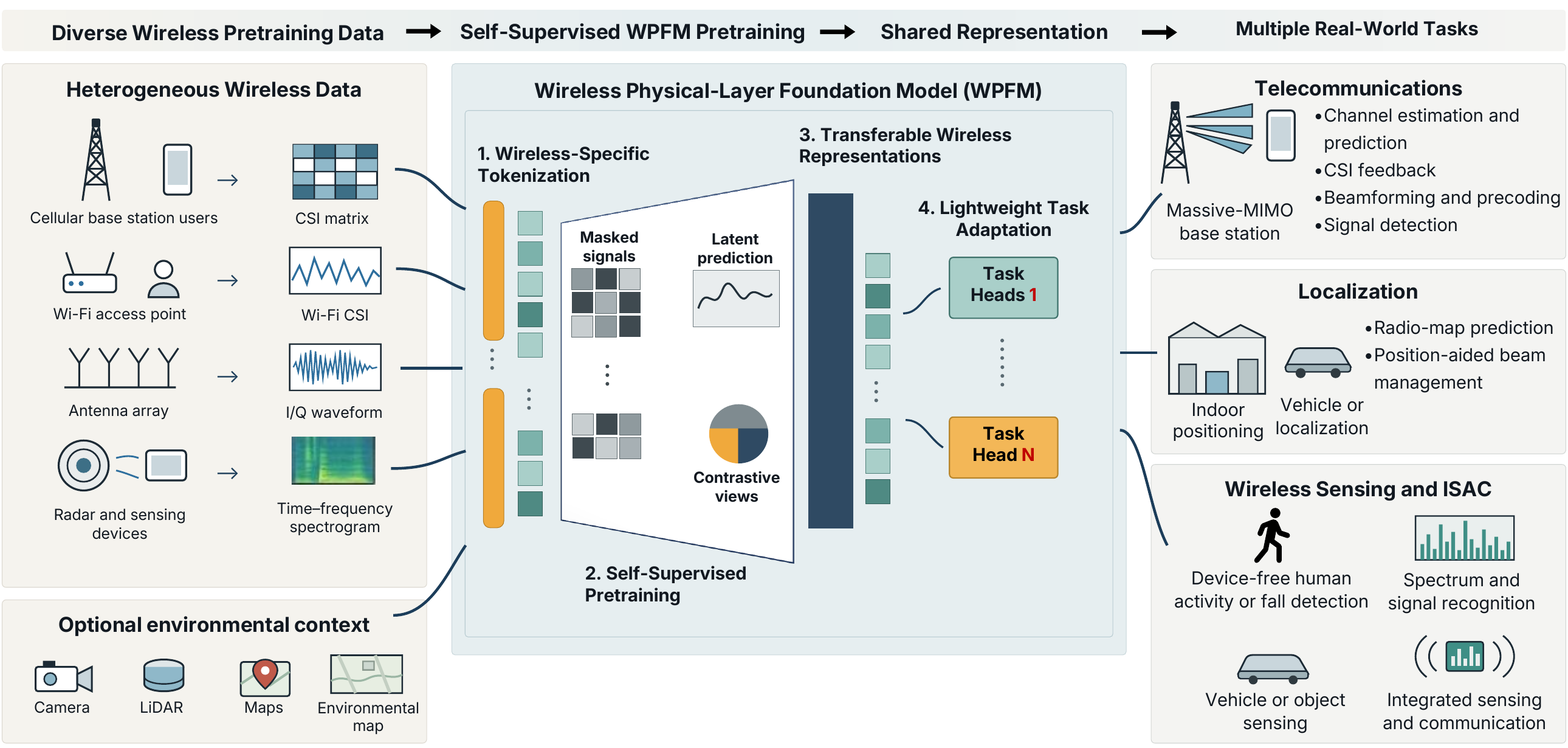}
    \caption{WPFM workflow. Heterogeneous wireless inputs are tokenized for self-supervised pretraining. The shared representation is adapted using task heads to support different applications.}

    \label{fig:wpfm_architecture}
\end{figure*}

\ac{ai} has become an important tool for wireless communication systems, addressing problems in channel estimation, signal detection, beamforming, positioning, and sensing that are difficult to solve with conventional model-based methods. However, most deployed wireless \ac{ai} solutions are task-specific, which means a separate model is designed, trained, and maintained for each task, dataset, and deployment configuration~\cite{9796036, 9134938}. This approach scales poorly as the number of tasks and configurations grows, and the resulting models generalize weakly when the frequency band, antenna array, or propagation environment changes. Fig.~\ref{fig:wpfm_architecture} illustrates the WPFM which is a shared backbone that learns representations from heterogeneous wireless data and is then adapted to multiple tasks.

A different paradigm has emerged in \ac{nlp} and computer vision, where large models are pretrained once on broad unlabeled data using \ac{ssl} and then adapted to many downstream tasks with little labeled data. Architectures such as the Transformer~\cite{vaswani2017attention}, together with pretraining methods such as masked language modeling~\cite{devlin2019bert} and masked image modeling~\cite{he2022masked}, allow a single backbone to transfer across tasks. These foundation models motivate a similar shift in wireless systems to move from many isolated task-specific models toward a shared, reusable backbone that can be adapted with limited data.

Transferring this approach to the wireless physical layer, however, is not straightforward. Unlike text, which is discrete and draws on a natural vocabulary, or images, which follow a standardized pixel grid, wireless signals are continuous, complex-valued, and high-dimensional. Their structure also varies with the system configuration, including the antenna count, bandwidth, and subcarrier spacing. Beyond these device-related factors, the propagation channel itself introduces noise, fading, and reflections. The statistics of wireless signals are therefore governed by propagation physics. In addition, these signals must be processed under strict latency constraints on hardware with limited memory and computational resources. As a result, general-purpose foundation models cannot be applied directly, and wireless-specific choices in tokenization, architecture, pretraining objective, and model size are required.

To address this gap, the concept of a \gls{wpfm}, a model pretrained on large amounts of unlabeled wireless data to learn transferable representations that support multiple physical-layer tasks, was articulated by Fontaine et al.~\cite{10615509}. Since then, a growing body of work has proposed concrete models, including LWM~\cite{alikhani2024largewirelessmodellwm} and WiFo~\cite{Liu2025}, spanning channel modeling, localization, and sensing. In parallel, industry and standardization bodies pursued in \ac{ai}-native air interfaces, with 3GPP Release-18 and Release-19 studying AI/ML-assisted \ac{csi} feedback, beam management, and positioning~\cite{lin2023overview3gppstudyartificial}. Although these efforts remain task-specific and do not yet target foundation models directly, their emphasis on cross-configuration generalization and reusable model management aligns with the \gls{wpfm} vision. More broadly, foundation models are increasingly discussed in both the academic and industrial literature as a candidate direction for 6G. This convergence of research and standardization activity motivates a dedicated review of the area.

This paper focuses specifically on the physical layer and reviews \glspl{wpfm} across their learning paradigms, architectures, applications, and deployment constraints. In contrast to prior surveys, which we discuss in the remainder of this section, we jointly treat pretraining paradigms, neural architectures, and fine-tuning strategies; organize existing models under a unified taxonomy; and analyze deployment feasibility by mapping model size to hardware constraints across the \ac{ue}--gNB--edge--cloud hierarchy. We cover telecommunications, localization, and sensing within a single framework, and dedicate attention to model compression as a bridge between large-scale models and resource-constrained wireless deployments.

\begin{figure*}
    \centering
    \includegraphics[width=\linewidth]{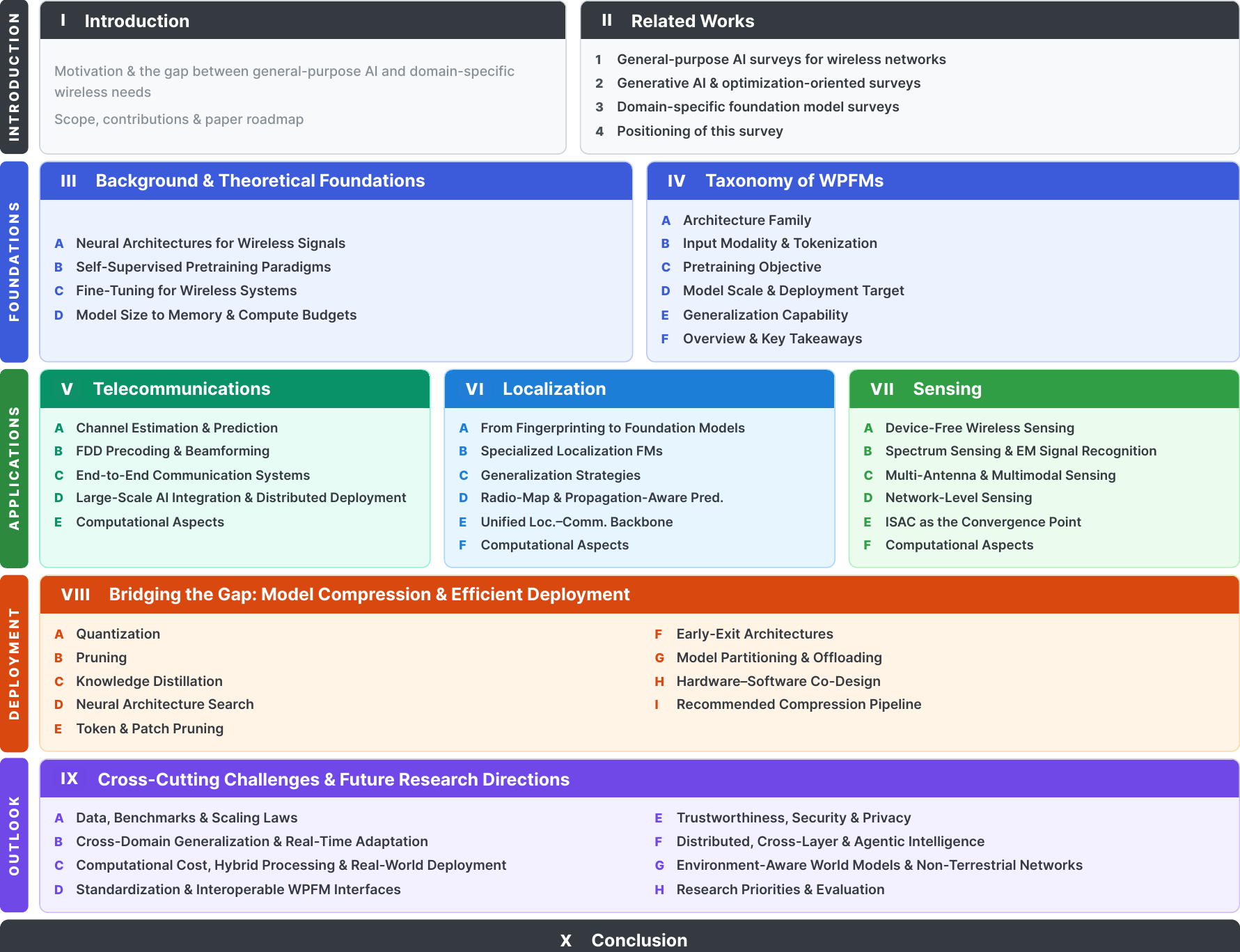}
    \caption{Structure of this survey, from theoretical foundations and the WPFM taxonomy to application domains, deployment considerations, and open research directions.}

    \label{fig:paper_architecture}
\end{figure*}

The remainder of this paper is organized as follows. Fig.~\ref{fig:paper_architecture} provides a visual overview of the paper structure. The acronyms used throughout the paper are summarized in Table~\ref{tab:acronyms}. Section~\ref{subsec:related_works} reviews related surveys and positions this work relative to them. Section~\ref{sec:theory} presents the theoretical background, covering neural architectures, self-supervised pretraining paradigms, fine-tuning strategies, and a framework for relating model size to memory and compute budgets. Section~\ref{sec:taxonomy} introduces a taxonomy of existing \glspl{wpfm}. Sections~\ref{sec:telecom}, \ref{sec:localization}, and \ref{sec:sensing} review applications in telecommunications, localization, and sensing, respectively, each including an analysis of computational feasibility. Section~\ref{sec:compression} discusses model compression and efficient deployment. Section~\ref{sec:challenges} examines cross-cutting challenges and the corresponding future research directions. Finally, Section~\ref{sec:conclusion} concludes the paper.

\begin{table*}[htbp]
\caption{Acronyms Used Throughout This Survey}
\label{tab:acronyms}
\centering
\renewcommand{\arraystretch}{0.9}
\footnotesize
\begin{tabular}{@{}l p{6.6cm} l p{6.5cm}@{}}
\toprule
\textbf{Acronym} & \textbf{Definition} & \textbf{Acronym} & \textbf{Definition} \\
\midrule
3GPP & Third Generation Partnership Project & MIMO & Multiple-Input Multiple-Output \\
AGV & Automated Guided Vehicle & MLP & Multi-Layer Perceptron \\
AI & Artificial Intelligence & MoE & Mixture of Experts \\
AirComp & Over-the-Air Computation & MSE & Mean Squared Error \\
AoA & Angle of Arrival & NAS & Neural Architecture Search \\
ASIC & Application-Specific Integrated Circuit & NLOS & Non-Line-of-Sight \\
AWGN & Additive White Gaussian Noise & NLP & Natural Language Processing \\
BBU & Baseband Unit & NPU & Neural Processing Unit \\
BERT & Bidirectional Encoder Representations from Transformers & NR & New Radio \\
BEV & Bird's Eye View & NTN & Non-Terrestrial Network \\
BFP & Block Floating Point & OFDM & Orthogonal Frequency Division Multiplexing \\
BIM & Basic Iterative Method & O-RAN & Open Radio Access Network \\
BLER & Block Error Rate & PEFT & Parameter-Efficient Fine-Tuning \\
BS & Base Station & PGD & Projected Gradient Descent \\
CFM & Channel Foundation Model & PRI & Pulse Repetition Interval \\
CIR & Channel Impulse Response & PTQ & Post-Training Quantization \\
CNN & Convolutional Neural Network & QAT & Quantization-Aware Training \\
CSI & Channel State Information & RAN & Radio Access Network \\
DMA & Direct Memory Access & ReLU & Rectified Linear Unit \\
DNN & Deep Neural Network & RF & Radio Frequency \\
DSP & Digital Signal Processor & RGB & Red, Green, and Blue \\
DTI & Domain-Transformation Invariance & RIC & RAN Intelligent Controller \\
EIC & Environment-Intelligent Communication & RMa & Rural Macrocell \\
FDD & Frequency Division Duplex & RMSE & Root Mean Squared Error \\
FFT & Fast Fourier Transform & RoPE & Rotary Positional Embedding \\
FGSM & Fast Gradient Sign Method & RNN & Recurrent Neural Network \\
FLOPs & Floating-Point Operations & RSRP & Reference Signal Received Power \\
FMCW & Frequency-Modulated Continuous Wave & S-R MoE & Shared-and-Routed Mixture of Experts \\
FP16/FP32 & 16-/32-Bit Floating Point & SFFN & Spiking Feed-Forward Network \\
FPGA & Field-Programmable Gate Array & SIGINT & Signals Intelligence \\
GLUE & General Language Understanding Evaluation & SNN & Spiking Neural Network \\
gNB & Next-Generation Node B & SNR & Signal-to-Noise Ratio \\
GNSS & Global Navigation Satellite System & SRAM & Static Random-Access Memory \\
GPS & Global Positioning System & SS-DMoE & Shared-Specific Disentangled Mixture of Experts \\
GPU & Graphics Processing Unit & SSL & Self-Supervised Learning \\
HAR & Human Activity Recognition & SSM & State Space Model \\
ICL & In-Context Learning & STF-PE & Space-Time-Frequency Positional Encoding \\
IEEE & Institute of Electrical and Electronics Engineers & STFT & Short-Time Fourier Transform \\
IMU & Inertial Measurement Unit & SWaP & Size, Weight, and Power \\
INT4/INT8 & 4-/8-Bit Integer & TDD & Time Division Duplex \\
IoT & Internet of Things & TOPS & Tera Operations per Second \\
IQ & In-phase and Quadrature & TTT & Test-Time Training \\
ISAC & Integrated Sensing and Communication & TWM & Telecom World Model \\
JEPA & Joint Embedding Predictive Architecture & UAP & Universal Adversarial Perturbation \\
KD & Knowledge Distillation & UAV & Unmanned Aerial Vehicle \\
LDPC & Low-Density Parity-Check & UE & User Equipment \\
LIF & Leaky Integrate-and-Fire & UMa & Urban Macrocell \\
LLM & Large Language Model & UMi & Urban Microcell \\
LMF & Location Management Function & URLLC & Ultra-Reliable Low-Latency Communications \\
LoRA & Low-Rank Adaptation & UWB & Ultra-Wideband \\
LOS & Line-of-Sight & ViT & Vision Transformer \\
MAC & Medium Access Control & WFM & Wireless Foundation Models \\
MAE & Masked Autoencoder; Mean Absolute Error & WPFM & Wireless Physical-Layer Foundation Model \\
MIM & Momentum Iterative Method & WWM & Wireless World Model \\
ML & Machine Learning & XR & Extended Reality \\
\bottomrule
\end{tabular}
\end{table*}

\begin{table*}[t]
\centering
\caption{Scope and Methodological Characteristics of Related Surveys on Foundation Models for Wireless Communications}
\label{tab:related_works}
\renewcommand{\arraystretch}{1.12}
\setlength{\tabcolsep}{3pt}
\scriptsize
\begin{tabularx}{\textwidth}{@{}p{0.95cm} p{2.1cm} p{3cm} p{1.2cm} p{2.3cm} p{3cm} X@{}}
\toprule
\textbf{Ref.} & \textbf{Declared Scope} & \textbf{Organizing Framework} & \textbf{PHY Treatment} & \textbf{Learning / Adaptation} & \textbf{Application Emphasis} & \textbf{Deployment Treatment}  \\
\midrule
\cite{11370829} & Large AI models for future communications & Eight-dimensional comparison of architectures, adaptability, modalities, optimization, knowledge, scope, deployment, and risks & Component & Architectures; pretraining; fine-tuning; alignment & PHY design, resource allocation, management, edge intelligence, semantic communication, and agents & Deployment is an explicit comparison axis; resource-constrained deployment is discussed qualitatively\\
\midrule
\cite{10683090} & Convergence of foundation models and 6G networks & FM fundamentals, 6G enabling mechanisms, and application scenarios & Adjacent & Pretraining overview; PEFT; federated device--edge fine-tuning & Edge computing, federated learning, blockchain, IoV, Metaverse, and security & MEC and federated device--edge frameworks  \\
\midrule
\cite{11316403} & Multimodal FMs for wireless prediction and control & FM/modality foundations followed by prediction and control task groups & Component & FM architectures; multimodal pretraining and fine-tuning & Traffic and channel prediction, blockage prediction, beam selection, handover, and control & Cloud, edge, on-device, and collaborative deployment paradigms  \\
\midrule
\cite{electronics15071345} & Impact of generative AI and LLMs on 6G architecture and services & Five inductive clusters: architecture, management, security, semantics, and edge AI & Component & GenAI architectures; adaptation and edge-efficiency methods & Full network stack from generative PHY and semantics to management, security, and edge AI & Split learning, on-device inference, and extreme quantization are examined as deployment issues \\
\midrule
\cite{zou2026telecomworldmodelsunifying} & Telecom world models combining digital twins, FMs, and planning & Five world-model paradigms and a two-world, three-layer TWM architecture & Component & Layer-specific training for field, dynamics/control, and telecom-FM layers & Field prediction, network control, slicing, contingency planning, and intent translation & Construction/evaluation methodology, proof-of-concept, and production-deployment challenges  \\
\midrule
\cite{jiang2025channelfoundationmodelscfms} & Channel foundation models & CFM methods grouped as generative, discriminative, and combined approaches & Dedicated & Masked channel modeling; contrastive learning; downstream adaptation & Channel/PHY tasks, RAN functions, and ISAC & Model advancement and efficiency are discussed as open challenges; no hardware comparison \\
\midrule
\cite{ZHOU2026} & Generative and large AI models for 6G optimization & Three optimization classes: information compression, beamforming, and mixed-integer satellite optimization & Adjacent & Generative-model and LAM principles; no WPFM adaptation taxonomy & Information transmission, cell-free/ISAC beamforming, and satellite microservice placement & Conceptual optimization frameworks, including a deployment-optimization case  \\
\bottomrule
\end{tabularx}
\vspace{1mm}

\raggedright\footnotesize{Entries describe explicitly documented "Dedicated" denotes a survey centered on physical-layer or channel foundation models; "Component" denotes an explicit physical-layer section or architectural layer within a broader communication survey; and "Adjacent" denotes a primarily network-level scope.}
\end{table*}

\section{Related Works}
\label{subsec:related_works}
Several recent surveys have examined the intersection of foundation models and wireless communications from various angles. In this subsection, we categorize them into three groups: general-purpose \ac{ai} surveys for wireless networks, generative \ac{ai} and optimization-oriented surveys, and domain-specific foundation model surveys. We review each group, identify the aspects they cover and the gaps they leave, and conclude by positioning our work relative to them. Table~\ref{tab:related_works} provides a structured comparison across key dimensions.

\subsubsection{General-Purpose AI Surveys for Wireless Networks}
Several surveys study the role of large \ac{ai} models in future wireless networks from a broad perspective. \cite{11370829} presents a survey of large \ac{ai} models for future communications, including \glspl{llm}, large vision models, large multimodal models, and world models. The survey examines diverse model architectures, such as Transformer~\cite{vaswani2017attention}, diffusion~\cite{ho2020denoising}, and Mamba~\cite{gu2024mambalineartimesequencemodeling}, and their applications across the entire network stack. However, it briefly addresses the physical layer as one of several application domains rather than as its central focus. The survey does not provide a detailed taxonomy of \glspl{wpfm}, and it does not analyze deployment feasibility in terms of model size versus hardware constraints.

On the other hand, \cite{10683090} explores the convergence of \ac{ai} foundation models with 6G wireless networks, identifying federated learning, mobile edge computing, and Metaverse applications as key deployment scenarios. This work provides a framing of the relationship between foundation models and 6G, but it remains at a high level without a detailed analysis of physical-layer pretraining objectives or architectural design choices specific to wireless signals.

\subsubsection{Generative AI and Optimization-Oriented Surveys}
\cite{ZHOU2026} proposes leveraging generative and large \ac{ai} models for scalable network optimization, focusing on information compression, beamforming design, and automated optimization for satellite networks. The survey provides insights into diffusion-based generation for multi-objective optimization, but it is narrowly focused on the optimization perspective and does not address \ac{ssl} pretraining paradigms, signal-level representation learning, or the broader range of physical-layer tasks (e.g., localization, sensing, spectrum management). Additionally, \cite{electronics15071345} conducts a literature review on the impact of generative \ac{ai} on 6G architecture, categorizing contributions into five clusters: architecture, management, security, semantics, and edge \ac{ai}. This survey does not specifically examine \glspl{wpfm}, their pretraining strategies or the computational feasibility of deploying such models on telecom hardware.

\subsubsection{Domain-Specific Foundation Model Surveys}
\cite{11316403} includes multi-modal data-enhanced foundation models for prediction and control in wireless networks. This work provides a review of how foundation models process multi-modal contextual information (visual, graph, point cloud, and RF data) and applies it to prediction and control tasks. It also compares cloud, edge, on-device, and collaborative deployment paradigms. However, it does not organize the literature around physical-layer pretraining objectives or provide a quantitative mapping from model requirements to wireless hardware budgets. On the other hand, \cite{jiang2025channelfoundationmodelscfms} introduces the concept of \glspl{cfm} and provides a review of \ac{ssl} methods for channel-related tasks such as channel estimation, channel prediction, \ac{csi} feedback and compression, and beam management, categorizing them into generative, discriminative, and combined paradigms. This work is the closest to ours in its physical-layer focus; however, \glspl{cfm} are restricted to channel-related tasks and do not provide a unified treatment of localization, sensing, and \ac{isac}. Furthermore, the survey does not systematically compare fine-tuning strategies, model-compression pipelines, or the computational constraints of deploying foundation models. Addressing fine-tuning strategies is crucial because the value of a pretrained \gls{wpfm} is ultimately realized during adaptation to diverse downstream tasks under scarce labeled data and shifting channel conditions; the choice of fine-tuning strategy (e.g., full fine-tuning versus parameter-efficient adaptation) directly impacts the accuracy, cost, and deployment feasibility on resource-constrained hardware.

To survey the world models, \cite{zou2026telecomworldmodelsunifying} proposes the \ac{twm} concept, a three-layer architecture that unifies digital twins, foundation models, and planning for 6G. The \ac{twm} framework introduces a field world model for physical-layer dynamics, a control/dynamics world model for network-level decisions, and a telecom foundation model layer for intent translation. While innovative in its systems-level perspective, this work is conceptual and focuses on world-model architectures rather than providing a comprehensive review of existing \glspl{wpfm} and their empirical performance.

\subsubsection{Positioning of This Survey}
In contrast to the surveys reviewed above, this paper is, to our knowledge, the first to jointly cover pretraining, architecture, fine-tuning, and deployment for \glspl{wpfm} across telecommunications, localization, and sensing. Our work makes several contributions that are not found in any single prior survey:

\begin{itemize}
    \item \textbf{Physical-layer focus with a unified taxonomy:} We classify \glspl{wpfm} in five dimensions, architecture family, input modality, pretraining objective, model scale/deployment target, and generalization capability, providing a review of the area that is absent from prior works.
    \item \textbf{Joint treatment of pretraining, architecture, and fine-tuning:} We provide an integrated analysis of \ac{ssl} pretraining paradigms, neural architectures, and fine-tuning strategies, whereas existing surveys typically address only a subset of these dimensions.
    \item \textbf{Computational feasibility analysis:} We provide a quantitative framework (Eqs.~\eqref{eq:weight_mem}--\eqref{eq:max_params}) for mapping model sizes to hardware constraints across the UE--gNB--edge--cloud hierarchy, and apply it to each application domain. No prior survey provides this level of deployment-aware analysis.
    \item \textbf{Unified cross-domain analysis:} Prior surveys focus on a single domain, typically channel-related communication tasks; we jointly review telecommunications, localization, and sensing and show how a shared pretrained backbone can serve all.
    \item \textbf{Model compression and efficient deployment:} We dedicate a full section to bridging the gap between large-scale foundation models and resource-constrained wireless deployments.
\end{itemize}

\section{Background and Theoretical Foundations}
\label{sec:theory}
 
In this section, we establish the necessary theoretical grounding to understand wireless foundation models. This section covers the neural architectures
most commonly employed, the core pretraining paradigms, and the fine-tuning strategies used to adapt pretrained models to downstream wireless tasks. The concept of a wireless \gls{wpfm} was formally articulated by Fontaine et al.~\cite{10615509}, who defined core challenges, such as building domain-specific pretraining, designing task-agnostic representation objectives, and devising adaptation strategies for heterogeneous wireless environments and proposed initial strategies to address them. This work established the foundation for the research in the next sections. To enhance the value of academic outputs and influence future foundation model studies, we address practical constraints related to model size, memory, and computational budgets.

\subsection{Neural Architectures for Wireless Signals}
\label{subsec:architectures}

Models can have different architectures based on the application and the requirements. We have summarized the models in Table~\ref{tab:arch_comparison}, and the next sections review the most used architectures.

\begin{table*}[ht]
\centering
\caption{Comparison of Neural Architecture Families Used in WPFMs. In the complexity column, $n$ denotes the input sequence length, i.e., the number of tokens.}

\label{tab:arch_comparison}
\renewcommand{\arraystretch}{1}
\resizebox{\textwidth}{!}
{%
\begin{tabular}{m{1.6cm}m{2.2cm} c m{4cm}m{4cm}m{2cm}}
\toprule
\textbf{Architecture} &
\textbf{Mechanism} &
\textbf{Complexity} &
\textbf{Strengths} &
\textbf{Limitations} &
\textbf{Wireless Inputs} \\
\midrule
Transformer &
Self-attention &
$\mathcal{O}(n^2)$ &
Long-range dependency capture; scalable to long sequences and multi-antenna arrays &
Quadratic cost on long sequences; requires task-specific tokenization for continuous RF signals &
IQ streams, CSI matrices, spectrograms \\
\cmidrule(lr){1-6}
Hybrid CNN-Transformer &
CNN front-end + Transformer backend &
$\mathcal{O}(n^2)$ &
Local feature extraction (Doppler shifts, subcarrier fading) combined with global temporal aggregation; exploits translation-invariant local patterns &
Increased architectural complexity; CNN and Transformer modules must be co-designed and jointly tuned &
IQ streams, spectrograms \\
\cmidrule(lr){1-6}
ViT Variants &
2-D patch-based self-attention &
$\mathcal{O}(n^2)$ &
Strong spatial feature learning; proven strategy for CSI/spectrogram inputs; task-specific MLP heads enable multitask learning &
Standard isotropic patches may blend independent physical dimensions (time, frequency, antenna); requires asymmetric patches or separable attention &
CSI matrices, spectrograms, radio quality maps \\
\cmidrule(lr){1-6}
SSMs &
Continuous-time differential equations (e.g., Mamba~\cite{gu2024mambalineartimesequencemodeling}) &
$\mathcal{O}(n)$ &
Linear-time sequence modeling; physics-aligned continuous-time formulation; reduced latency and memory footprint; edge-deployment friendly &
Less mature libraries compared to Transformers; limited adoption in wireless FM literature to date &
IQ signals, RF streams \\
\cmidrule(lr){1-6}
Multimodal Architectures &
Cross-modal attention / hierarchical temporal pooling &
$\mathcal{O}(n^2)$ &
Encodes heterogeneous inputs; improved generalization to unseen scenarios through modality fusion &
Extreme sampling rate disparity between RF and low-frequency modalities requires refined cross-modal alignment mechanisms &
IQ signals, CSI, text, images, geo-coordinates \\
\bottomrule
\end{tabular}%
}
\end{table*}

\subsubsection{Transformer-Based Architectures}
The Transformer~\cite{vaswani2017attention} replaced recurrence and convolution with a self-attention mechanism that relates every element of a sequence to every other element in parallel, and it has since become the dominant backbone across natural-language processing and vision. These sequence elements are called tokens: each token is a vector that represents a small unit of the input (for example, a word in text, an image patch in vision, or, in the wireless setting, a short segment of an \ac{iq} stream or a group of subcarriers of a \ac{csi} matrix), as illustrated in Fig.~\ref{fig:architecture_families}. In self-attention, each token is projected into three vectors, a query ($Q$), a key ($K$), and a value ($V$); the similarity between a token's query and every other token's key determines how much of each value is aggregated into that token's updated representation. Because every token is compared against every other one, self-attention can capture long-range dependencies irrespective of their distance in the sequence, an ability that is valuable for temporal wireless signals and multi-antenna arrays.

This all-pairs comparison, however, makes the computation and memory grow quadratically ($\mathcal{O}(n^2)$) with the sequence length $n$ (Table~\ref{tab:arch_comparison}), which becomes the dominant cost for long \ac{iq} or \ac{csi} sequences. Transformer-based designs have been adopted in LWM~\cite{alikhani2024largewirelessmodellwm},
Cheraghinia et al.~\cite{11264506}, and LWM-Spectro~\cite{kim2026lwmspectrofoundationmodelwireless}, while WirelessGPT~\cite{11278185} employs a hybrid CNN-Transformer architecture with convolutional multi-scale patch embedding. Unlike discrete text with a natural vocabulary, continuous IQ streams must first be tokenized via patching or learned convolutional embeddings to map noisy, high-dimensional physical signals into a latent space suitable for the attention mechanism.

\subsubsection{Hybrid CNN-Transformer Architectures}
\glspl{cnn}~\cite{lecun1998gradient} apply learnable convolutional filters with shared weights across the input, giving them a built-in bias toward local, translation-invariant patterns. This is advantageous for wireless data where local patterns, such as Doppler shifts or subcarrier fading, exhibit translation-invariant properties that \glspl{cnn} inherently capture it, letting the Transformer focus entirely on macroscopic temporal dynamics. \ac{cnn} front-ends can extract local time-frequency features, while Transformer backends aggregate global features. This hybrid design is well-suited to wireless inputs and has been explored in models such as WirelessGPT~\cite{11278185}.

\subsubsection{ViT Variants}
The \ac{vit}~\cite{dosovitskiy2021image} adapts the Transformer to images by splitting an input into a grid of fixed-size patches, linearly embedding each patch as a token, and processing the resulting sequence with standard self-attention. \gls{vit}-based models demonstrate spatial capabilities in both the image domain and the wireless domain. Treating \ac{csi} matrices or spectrograms as images with two-dimensional patches is a proven and tested strategy. Nevertheless, because the axes of wireless data (e.g., time, frequency, and antenna space) represent distinct physical domains unlike the isotropic axes of standard RGB images, wireless-specific \glspl{vit} frequently employ asymmetric patch sizes or separable attention mechanisms. This prevents the blending of independent physical dimensions and preserves the integrity of the channel structure. MapViT~\cite{hsu2026mapvittwostagevitbasedframework}
applies a two-stage \ac{vit} framework to radio quality map prediction.
6G WavesFM~\cite{11131142} uses a shared \ac{vit} backbone with task-specific \ac{mlp} heads supporting a wide array of communication, sensing, and localization tasks.

\subsubsection{State-Space Model Architectures}
As an alternative to the quadratic complexity of Transformer attention,
\glspl{ssm}~\cite{gu2022efficiently} provide linear-time sequence modeling. \Glspl{ssm} map an input sequence to an output through a latent state governed by linear recurrence, which can be evaluated as a global convolution and scales linearly with sequence length; Mamba~\cite{gu2024mambalineartimesequencemodeling} extends this formulation with input-dependent (selective) state transitions. Beyond their computational efficiency, \glspl{ssm} are derived from continuous-time differential equations. This continuous-time mathematical foundation aligns naturally with the physics of analog radio-frequency signals, allowing models like WiMamba~\cite{raviv2026wimambalinearscalewirelessfoundation} to capture signal dynamics better than architectures constrained by discrete tokenization. WiMamba builds upon the Mamba \cite{gu2024mambalineartimesequencemodeling} architecture, replacing attention mechanisms with a \ac{ssm} and combining this with a preprocessing module and a self-supervised pretraining framework. WiMamba matches or outperforms Transformer-based wireless foundation models across four downstream tasks while delivering significant reductions in latency and memory footprint, making it a particularly compelling option for edge and resource-constrained deployments.

\subsubsection{Multimodal Architectures}
Multimodal wireless foundation models encode heterogeneous inputs, such as \ac{iq} signals, \ac{csi}, text descriptions, and geographic coordinates. Studies \cite{10599304, aboulfotouh2026multimodalwirelessfoundationmodels, 10971878} motivate multimodal approaches, with Jiao et al.~\cite{10971878} demonstrating that combining modalities can improve generalization to unseen scenarios. A critical challenge in this domain is the extreme disparity in sampling rates; aligning MHz-scale \ac{rf} streams with low-frequency text or camera feeds requires refined cross-modal attention or hierarchical temporal pooling mechanisms to synchronize latent representations without ignoring high-resolution wireless dynamics.

\subsubsection{Trade-offs and Comparison} Selecting an appropriate neural architecture for wireless foundation models requires balancing representation capacity against the physical constraints of deployment.

\begin{itemize}
\item \textbf{Pure Transformers} capture global context across antenna arrays and long temporal windows through all-pairs self-attention, but their $\mathcal{O}(n^2)$ complexity increases latency and memory cost, which can be a limiting factor for real-time edge inference.
\item \textbf{Hybrid \ac{cnn}-Transformers and \glspl{vit}} combine convolutional or patch-based inductive biases with self-attention, which suits structured 2D inputs such as \ac{csi} matrices and spectrograms; their reliance on fixed patch grids is, however, less naturally matched to the raw continuous nature of 1D \ac{iq} streams.
\item  \textbf{\glspl{ssm}} offer long-context modeling at linear-time complexity, and their continuous-time formulation aligns with the physics of \ac{rf} signals; their main drawback is a less mature tooling ecosystem and, to date, more limited empirical validation in the wireless domain compared to Transformers.
\item \textbf{Multimodal architectures} can fuse heterogeneous inputs and are relevant as networks move toward semantic communication~\cite{xie2021deep} (which transmits the task-relevant meaning of a message rather than its exact bits) and \ac{isac}~\cite{kaushik2023integratedsensingcommunications6g} (which merges radar-like sensing and data communication onto shared spectrum and hardware), at the cost of higher model complexity and the need to align modalities with very different sampling rates.
\end{itemize}

The most suitable choice therefore depends on the target task and deployment point rather than on any single architecture dominating across the board.

\subsection{Self-Supervised Pretraining Paradigms}
\label{subsec:ssl}

\begin{figure*}
    \centering
    \includegraphics[width=\linewidth]{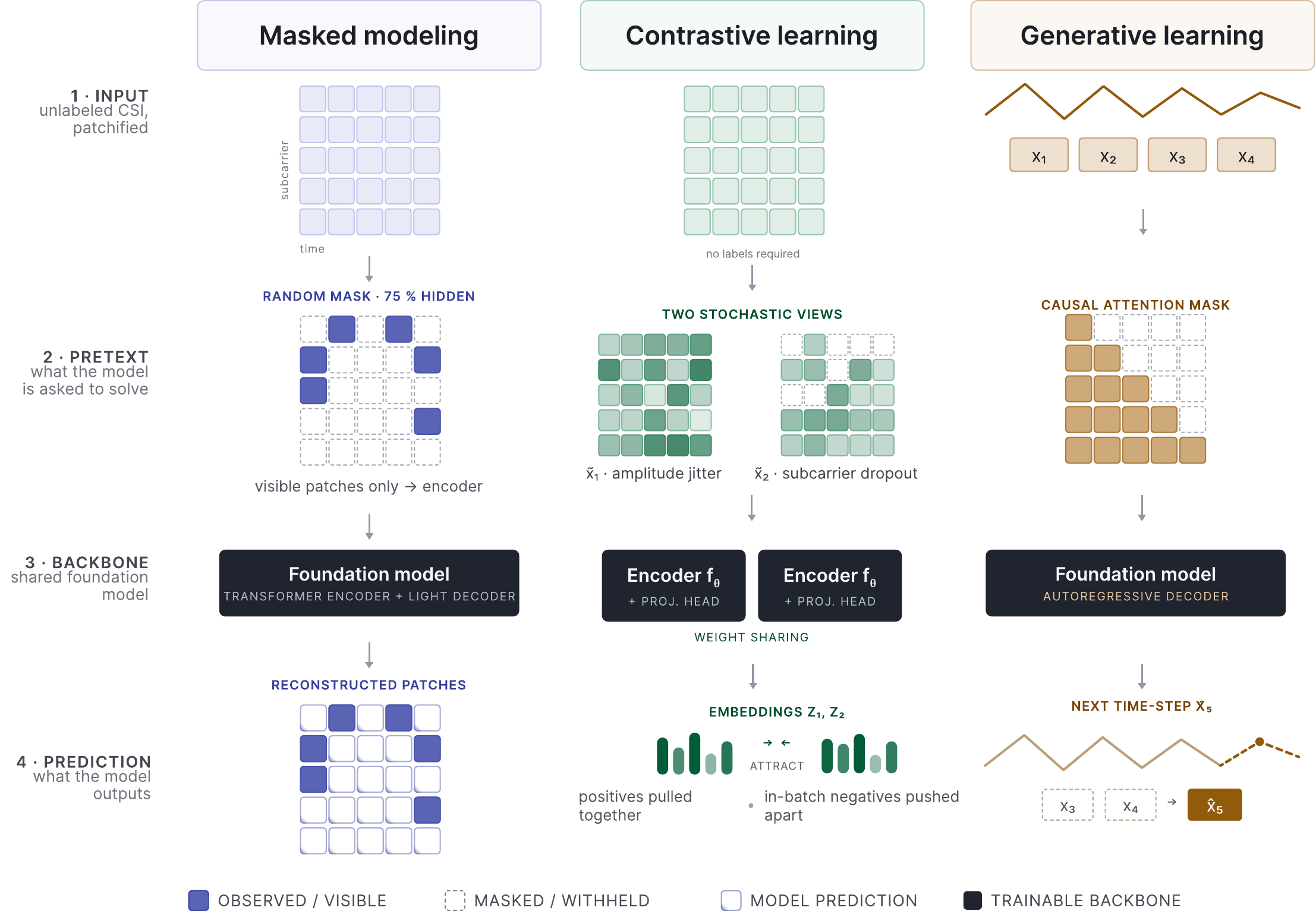}
    \caption{Self-supervised pretraining paradigms for WPFMs: (a) masked modeling reconstructs hidden signal patches, (b) contrastive learning aligns representations of augmented views, and (c) generative pretraining predicts future tokens or denoises corrupted signals.}
    \label{fig:ssl_methods}
\end{figure*}

\ac{ssl}~\cite{balestriero2023cookbook} eliminates the need for large labeled datasets by constructing supervisory signals from unlabeled data. Three main families of \ac{ssl} objectives used for wireless signal modeling are depicted in Fig.~\ref{fig:ssl_methods}.

\subsubsection{Masked Modeling}
Inspired by BERT~\cite{devlin2019bert} in \ac{nlp}, which pretrains a bidirectional Transformer by predicting randomly masked words, and the Masked Autoencoder~\cite{he2022masked} in computer vision, which reconstructs masked image patches from a small visible subset, masking a subset of the input (e.g., segments of an \ac{iq} stream or subcarriers of a \ac{csi} matrix) and trains the model to reconstruct the missing parts, forcing it to learn representations of the signal structure. Given an input channel matrix $\mathbf{H}$, a random binary mask $\mathcal{M}$ selects a subset of patches (or tokens) to be hidden. An encoder $f_\theta$ processes only the visible patches $\mathbf{H}_{\text{vis}}$, and a decoder $g_\phi$ reconstructs the masked portions. Unlike discrete text tokens in \ac{nlp}, wireless signals are continuous in nature. Consequently, most wireless masked modeling approaches employ a regression-based reconstruction loss, typically the \ac{mse} computed exclusively over the masked positions:
\begin{equation}
\label{eq:masked_loss}
\mathcal{L}_{\text{recon}} = \frac{1}{|\mathcal{M}|} \sum_{i \in \mathcal{M}} \left\| \mathbf{h}_i - \hat{\mathbf{h}}_i \right\|^2,
\end{equation}
where $\mathbf{h}_i$ and $\hat{\mathbf{h}}_i$ denote the original and reconstructed patches, respectively, and $|\mathcal{M}|$ is the number of masked patches. Restricting the loss to the masked subset is critical: it prevents the model from learning a trivial identity mapping and instead forces it to infer global spatial-frequency correlations from partial observations.

\subsubsection{Contrastive Learning}
Contrastive \ac{ssl} learns representations by pulling together embeddings of augmented views of the same sample (positive pairs) while pushing apart embeddings of different samples (negative pairs). The dominant loss function in this family is the InfoNCE objective~\cite{oord2018representation}, a categorical cross-entropy loss that identifies the true positive among a set of negative samples and thereby maximizes a lower bound on the mutual information between the two views. Given a batch of $B$ samples, each sample is augmented twice to produce $2B$ views. Let $\mathbf{z}_k$ denote the normalized embedding of the $k$-th view, and let $\text{pos}(k)$ be the index of the other view generated from the same source sample, so that $\mathbf{z}_{\text{pos}(k)}$ forms the positive pair of $\mathbf{z}_k$. The contrastive loss is:

\begin{equation}
\label{eq:contrastive_loss}
\mathcal{L}_{\text{contra}} = -\frac{1}{2B} \sum_{k=1}^{2B} \log \frac{\exp\!\left(\mathbf{z}_k \cdot \mathbf{z}_{\text{pos}(k)} / \tau\right)}{\sum_{j=1, j \neq k}^{2B} \exp\!\left(\mathbf{z}_k \cdot \mathbf{z}_j / \tau\right)},
\end{equation}

\noindent where $\tau > 0$ is a temperature parameter that controls the sharpness of the distribution. A lower $\tau$ amplifies discrimination between positive and negative pairs, while a higher $\tau$ produces softer probability distributions. The denominator implicitly treats all other $2B-2$ samples in the batch as negatives, which means that the quality of learned representations is sensitive to the batch size and the choice of augmentation strategy.

However, contrastive methods face a practical challenge in the wireless domain: temporally adjacent channel snapshots are often highly correlated, increasing the risk of false negatives, treating genuinely similar channels from nearby time slots as negative pairs. This can degrade the learned representations. To mitigate this, non-contrastive self-distillation approaches~\cite{caron2021emerging}, in which a network learns by matching its own predictions to those of a slowly evolving copy of itself rather than contrasting against negatives, are increasingly adopted; these rely on momentum-updated teacher networks rather than explicit negative pairs. In such frameworks, a student network $f_\theta$ is trained to match the output of a slowly evolving teacher $f_\xi$, whose weights are updated as an exponential moving average $\xi \leftarrow \lambda \xi + (1 - \lambda) \theta$, with $\lambda$ close to 1. The asymmetry between the teacher and student, combined with techniques such as centering and sharpening of the teacher output, prevents representation collapse without requiring negative pairs~\cite{katsuki2026tsdino}.
 
\subsubsection{Generative Pretraining}
Autoregressive and diffusion-based generative models learn the data distribution by predicting future tokens or denoising corrupted inputs.

\textit{Autoregressive models} decompose the joint distribution of a sequence of tokens $\mathbf{x} = (x_1, x_2, \ldots, x_T)$ via the chain rule of probability:
\begin{equation}
\label{eq:autoregressive}
p(\mathbf{x}) = \prod_{t=1}^{T} p(x_t \mid x_1, \ldots, x_{t-1}),
\end{equation}
and minimize the negative log-likelihood $\mathcal{L}_{\text{AR}} = -\sum_{t=1}^{T} \log p_\theta(x_t \mid x_{<t})$. This factorization naturally aligns with the sequential nature of temporal \ac{iq} streams.

\textit{Diffusion models}~\cite{ho2020denoising} take an alternative approach by progressively corrupting data with Gaussian noise over $T$ steps and training a network $\boldsymbol{\epsilon}_\theta$ to reverse the process, learning to generate samples by iteratively denoising from pure noise. The training objective simplifies to a denoising score-matching loss:
\begin{equation}
\label{eq:diffusion}
\mathcal{L}_{\text{diff}} = \mathbb{E}_{t, \mathbf{x}_0, \boldsymbol{\epsilon}} \left[ \left\| \boldsymbol{\epsilon} - \boldsymbol{\epsilon}_\theta(\mathbf{x}_t, t) \right\|^2 \right],
\end{equation}
where $\mathbf{x}_t$ is the noised sample at step $t$ and $\boldsymbol{\epsilon} \sim \mathcal{N}(\mathbf{0}, \mathbf{I})$ is the injected noise. Diffusion models excel at modeling complex spatial distributions, making them particularly suited for generating high-fidelity \ac{csi} matrices for \ac{mimo} systems where the joint distribution over antennas and subcarriers has rich multimodal structure. The potential of large generative \ac{ai} for the telecoms industry was emphasized by~\cite{10384630}, who argued that such models represent the next paradigm shift in network intelligence.


\begin{table*}[t]
    \centering
    \caption{Comparison of WPFM Adaptation Strategies by Overhead, Data Requirement, Performance Potential, and Deployment Target}

    \begin{tabular}{l|c|c|c|p{8.5cm}}
        \toprule
        \textbf{Fine-Tuning Method} & \textbf{Overhead} & \textbf{Sample Req.} & \textbf{Perf. Bound} & \textbf{Primary Target \& Use Case} \\
        \midrule
        
        \multirow{2}{*}{\textbf{Linear Probing}} & \multirow{2}{*}{low} & \multirow{2}{*}{low} & \multirow{2}{*}{low} & \textbf{Target:} Edge Device. \\
        & & & & \textbf{Use Case:} Low compute adaptation on edge devices; restricted tasks; quantifying embedding quality. \\
        \midrule
        
        \multirow{2}{*}{\textbf{Full Fine-Tuning}} & \multirow{2}{*}{High} & \multirow{2}{*}{High} & \multirow{2}{*}{High} & \textbf{Target:} Network Core / Cloud Server. \\
        & & & & \textbf{Use Case:} Max performance for complex non-linear tasks (e.g., massive MIMO); high expressivity. \\
        \midrule
        
        \multirow{2}{*}{\textbf{PEFT (e.g., LoRA)}} & \multirow{2}{*}{low} & \multirow{2}{*}{low} & \multirow{2}{*}{High} & \textbf{Target:} O-RAN Architecture. \\
        & & & & \textbf{Use Case:} Quick adaptation on base stations with modular updates; efficient multi-environment deployments. \\
        \midrule
        
        \textbf{Knowledge} & \multirow{2}{*}{Low} & \multirow{2}{*}{None} & \multirow{2}{*}{Low} & \textbf{Target:} Transferring model to extreme edge devices. \\
        \textbf{Distillation (KD)} & & & & \textbf{Use Case:} Addresses inability of UEs (e.g., IoT) to host massive models. \\
        \midrule
        
        \multirow{2}{*}{\textbf{Zero/Few-Shot}} & No & No & \multirow{2}{*}{High} & \textbf{Target:} Instant adaptation. \\
        & Retrain & Retrain  & & \textbf{Use Case:} Generalization to novel tasks/environments with zero retraining. \\
        \bottomrule
    \end{tabular}
    \label{tab:ft_comparison}
\end{table*}
\subsection{Fine-Tuning for Wireless Systems}
\label{subsec:transfer}

Effective transfer of pretrained representations to target wireless tasks requires careful adaptation methods. We have summarized the available methods with details in Table~\ref{tab:ft_comparison}. In the following sections, we go through the methods:

\subsubsection{Linear Probing} 
Linear probing~\cite{alain2017understanding} means freezing the entire pretrained foundation model and training only a single, linear layer (classifier or regressor) on top of the representations. Because only a linear map is learned, the resulting accuracy directly reflects how linearly separable, and therefore how informative, the frozen representations are. In the context of wireless systems, this paradigm serves two purposes: it is an evaluation metric to quantify the separability and quality of the frozen \ac{ssl} embeddings, and it provides a low-compute adaptation solution for edge devices where backpropagation through a massive network is infeasible. However, its performance is fundamentally bottlenecked if the downstream wireless task, such as complex multi-user beamforming, requires non-linear transformations not captured by the frozen backbone.

\subsubsection{Full Fine-Tuning}
In this method all model parameters are updated on labeled downstream data. While most expressive, requires significant compute and labeled samples, limiting its applicability at the wireless edge. Furthermore, full fine-tuning poses a high risk of catastrophic forgetting. As the model heavily adapts to the specific channel statistics of a localized deployment, it may overwrite the generalized, cross-domain propagation physics learned during pretraining.

 \subsubsection{Parameter-Efficient Fine-Tuning}
Techniques such as \ac{lora}~\cite{hu2022lora}, adapters, and prompt tuning freeze most pretrained parameters and introduce a small number of trainable modules. In particular, \ac{lora} injects trainable low-rank matrices into each frozen weight matrix, so that adapting a model requires updating and storing only a tiny fraction of its parameters. Device-edge cooperative fine-tuning~\cite{10558820} proposes splitting this process between mobile devices and edge servers, making \ac{peft} practical in 6G deployments. From a network architecture perspective, \ac{peft} helps scalable multi-environment deployments. A single, frozen foundation model can reside at a base station, while dynamic, environment-specific \ac{lora} adapters are hot-swapped into memory depending on the specific user, cell sector, or frequency band being served.

\subsubsection{Knowledge Distillation}
Deploying billion-parameter foundation models directly onto resource-constrained \ac{ue} or IoT devices is practically impossible due to memory, latency, and power limitations. \ac{kd}~\cite{hinton2015distilling} addresses this by training a lightweight "student" model to mimic the soft outputs, intermediate feature maps, or attention matrices of the pretrained "teacher" foundation model, transferring the teacher's learned behavior into a far smaller network. In wireless networks, this allows the network core or edge server to maintain the heavy foundation model while transmitting compressed, task-specific student models to end users.

\subsubsection{Zero-Shot and Few-Shot Generalization}
The ultimate goal of a wireless foundation model is to generalize to new environments, frequency bands, and hardware configurations without retraining. In zero-shot generalization, a pretrained model is applied directly to a task or scenario it never saw during training, using no labeled examples and no parameter updates; in few-shot generalization, the model is given only a small handful of labeled examples of the new task to adapt from, in contrast to conventional training that requires large task-specific datasets. HeterCSI~\cite{zhang2026hetercsichanneladaptiveheterogeneouscsi} addresses heterogeneous \ac{csi} pretraining to enable cross-scale and cross-scenario generalization, and \cite{alikhani2024largewirelessmodellwm} evaluates zero-shot channel downstream tasks across diverse propagation scenarios. An ongoing challenge for zero-shot adaptation in this domain is the absence of a discrete, natural vocabulary. Unlike \ac{nlp}, where prompts are explicitly defined, 'prompting' a wireless foundation model requires innovative techniques, such as injecting pilot signals, geographic coordinates, or synthetic channel responses to guide the model's zero-shot inference.

\subsubsection{Trade-offs in Fine-tuning Strategies}

Selecting the optimal tuning strategy requires balancing performance, computational overhead, and deployment location across the wireless network, as compared in detail in Table~\ref{tab:ft_comparison}. In brief, linear probing has the lowest cost but suits only simpler tasks and depends heavily on pretraining quality, whereas full fine-tuning attains the highest accuracy on complex non-linear tasks (e.g., massive MIMO channel estimation) at the price of being confined to cloud or high-compute edge servers. Between these extremes, \ac{peft} such as \ac{lora} offers a favourable balance for O-RAN deployments by adapting a shared frozen backbone through lightweight, modular updates, while knowledge distillation is the most viable route to the extreme edge (IoT devices and \glspl{ue}) under strict latency and power budgets. Zero/few-shot generalization is the most attractive but least mature option, enabling instant adaptation to novel bands or geographies yet remaining highly dependent on the scale, diversity, and cross-modal alignment of the pretraining data.


\subsection{Model Size to Memory and Compute Budgets}
\label{subsec:sizing_guidelines}

It is essential to establish a unified analytical framework for evaluating whether a neural network model can be accommodated within a target hardware environment. The two primary limiting factors for neural network inference are \emph{memory capacity} (the ability to store model parameters and intermediate activations) and \emph{memory bandwidth or compute throughput} (the ability to complete inference within the required latency budget).

\subsubsection{Weight Memory Requirements}

The memory required to store model parameters is determined directly by the parameter count $N$ and numerical precision $b$ (bits per parameter):

\begin{equation}
\label{eq:weight_mem}
M_{\mathrm{weights}} = \frac{N \times b}{8} \;\; \text{[bytes]}.
\end{equation}

 For example, a model with 10 million parameters requires approximately 40\,MB in FP32, 10\,MB in INT8, and 5\,MB in INT4, significantly improving feasibility for FPGA-class hardware deployment.

\subsubsection{Activation and Workspace Memory}

There are differences between the memory requirements of inference and training. During \emph{training}, all intermediate activations must be retained for back-propagation, resulting in a total memory footprint approximately four times the weight memory in FP32 (including weights, gradients, and two Adam optimizer states), or roughly three times in mixed-precision training~\cite{micikevicius2018mixed, rajbhandari2020zero}. Activation checkpointing can further trade compute for memory by recomputing selected activations during back-propagation~\cite{korthikanti2023reducing}.

During \emph{inference}, there is no backward pass and therefore no need to store all intermediate activations simultaneously. A well-optimized inference engine only retains: (i) the current layer’s input and output (with the output serving as the next layer’s input while previously allocated buffers are released), and (ii) layer-specific temporary variables (e.g., $Q$, $K$, $V$ projections, attention score matrices, and feed-forward intermediates).

For a Transformer model with hidden dimension $d$, sequence length $S$, and batch size $B$, the peak activation memory in a single layer is typically $\sim10\, B \cdot S \cdot d$ elements. In wireless tasks (e.g., $B=1$, $S \approx 100$, $d=256$), this equates to approximately $256\,\mathrm{K}$ elements (roughly 1\,MB in FP32), which is only about 5\% of the weight memory of a 6-layer model with the same dimensions~\cite{ivanov2021data}.

Adding a modest allocation for runtime workspace (layer/operator temporaries, I/O staging, framework overhead), a practical rule for inference memory in wireless settings is:

\begin{equation}
\label{eq:total_mem}
M_{\mathrm{inference}} \approx (1.1\text{--}1.25) \times M_{\mathrm{weights}}
\end{equation}

That is, weight storage constitutes the dominant term for inference, with activations and workspace contributing an additional 10--25\%. This stands in stark contrast to training scenarios, where the multiplier can be $3$--$4\times$.

\subsubsection{Compute-Bound Latency Estimation}

Given a hardware accelerator capable of $T$~TOPS (tera operations per second), the minimum latency to process a single input through a model of $N$ parameters is approximately~\cite{williams2009roofline}:

\begin{equation}
\label{eq:compute_latency}
t_{\mathrm{compute}} \approx \frac{2N}{T \times 10^{12}} \;\; \text{[seconds]},
\end{equation}


\begin{figure}
    \centering
    \includegraphics[width=1.02\linewidth]{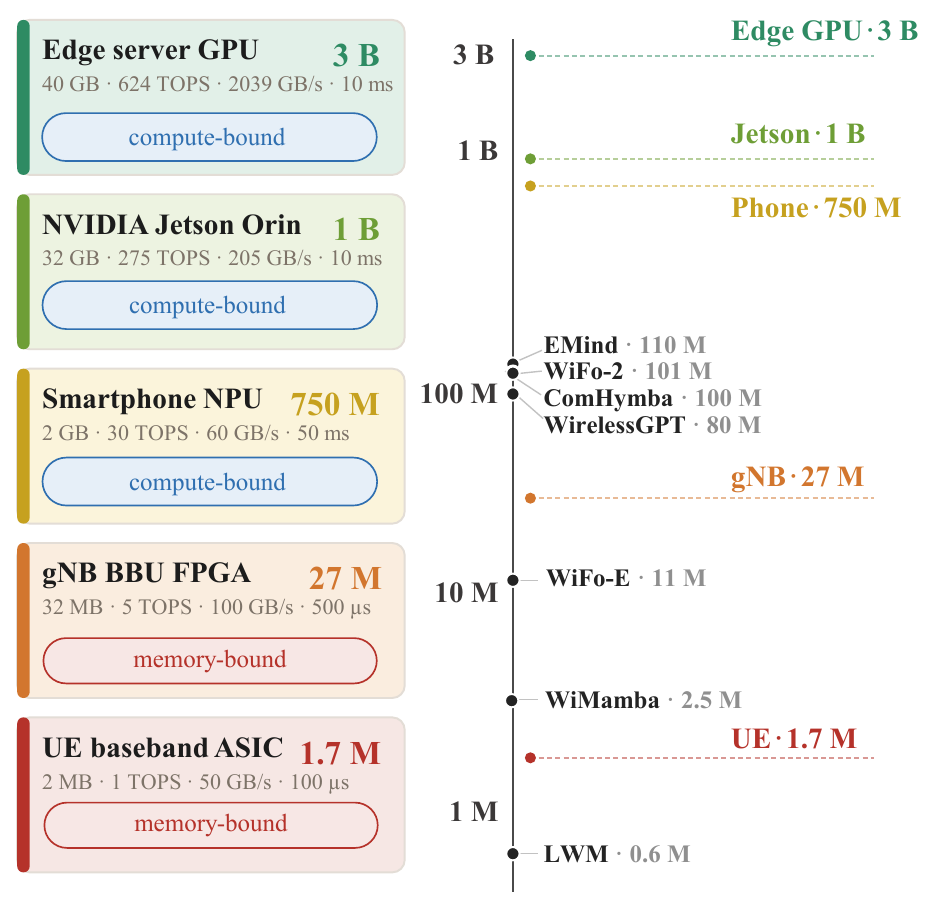}
    \caption{INT8 model-size limits for representative wireless platforms under memory, compute, bandwidth, and latency constraints, compared with selected WPFM parameter counts. UE and gNB platforms are memory-bound, whereas smartphone and edge-GPU platforms are compute-bound under the stated assumptions.}
    \label{fig:deployment_feasibility}
\end{figure}

where the factor of 2 reflects the multiply-accumulate operations prevalent in linear and attention layers. For example, a 5\,M-parameter model on a 1\,TOPS DSP yields a compute latency of approximately $10\,\mu\mathrm{s}$, while a 50\,M-parameter model on a 30\,TOPS NPU requires around $3.3\,\mathrm{ms}$.

\subsubsection{Memory Bandwidth-Bound Latency Estimation}

In scenarios with low batch sizes (common for real-time wireless inference), latency may be dominated by the time required to load model weights from memory rather than computational throughput. With memory bandwidth $\mathcal{B}$ (GB/s):

\begin{equation}
\label{eq:bw_latency}
t_{\mathrm{bandwidth}} \approx \frac{M_{\mathrm{weights}}}{\mathcal{B}}
\end{equation}

For instance, an FPGA with $50\,\mathrm{GB/s}$ bandwidth can load a $5\,\mathrm{MB}$ INT8 model in $100\,\mu\mathrm{s}$; a server GPU with $2\,\mathrm{TB/s}$ high-bandwidth memory loads a $200\,\mathrm{MB}$ model in a similar timeframe. The effective inference latency is $\max(t_{\mathrm{compute}},\, t_{\mathrm{bandwidth}})$.

\subsubsection{Practical Model Size Estimation and Resource Allocation}

Synthesizing the above equations, for a given hardware memory $M$, compute performance $T$, memory bandwidth $\mathcal{B}$, and latency budget $\Delta t$, the maximum deployable model parameter count is:

\begin{equation}
\label{eq:max_params}
N_{\max} = \min\!\left(
\frac{M}{1.2 \times (b/8)},\;\;
\frac{T \times 10^{12} \times \Delta t}{2},\;\;
\frac{\mathcal{B} \times \Delta t}{b/8}
\right)
\end{equation}

The factor $1.2$ in the memory term accounts for additional activation and workspace memory during inference (Eq.~\eqref{eq:total_mem}). Fig.~\ref{fig:deployment_feasibility} compares the resulting theoretical limits for representative platforms with selected WPFM parameter counts. Subsequent sections apply this formula to task-specific hardware and latency constraints.

Note that this formula presents theoretical upper bounds, assuming all resources are dedicated to the neural model. In practice, on-chip memory in ASICs and FPGAs is shared with other modules (e.g., FFT engines, LDPC decoders, DMA buffers, control logic), and only a fraction (10--30\%) is available for neural inference. Computation paths are also time-shared with conventional DSP functions. Therefore, practical model size estimates are purposefully more conservative. For platforms where the neural model has a dedicated accelerator (e.g., smartphone NPUs, edge GPUs), the gap between theoretical and practical limits is proportionally smaller.

\section{Taxonomy of Wireless Physical-Layer Foundation Models}
\label{sec:taxonomy}

While the previous section established the general theoretical foundations, including pretraining paradigms, neural architectures, and fine-tuning methods, this section shifts focus to \textbf{how} these building blocks are concretely combined and adapted in existing \glspl{wpfm}. We organize models along five groups: (i)~architecture family, (ii)~input modality and tokenization, (iii)~pretraining objective, (iv)~model scale and deployment target, and (v)~generalization capability. Table~\ref{tab:comparison} provides a detailed comparison of models across these dimensions. Throughout, we draw on evidence from the surveyed works to explain \textbf{why} specific design choices outperform alternatives in the wireless domain.

\subsection{Architecture Family}
\label{subsec:tax_arch}

\begin{figure*}
    \centering
    \includegraphics[width=\linewidth]{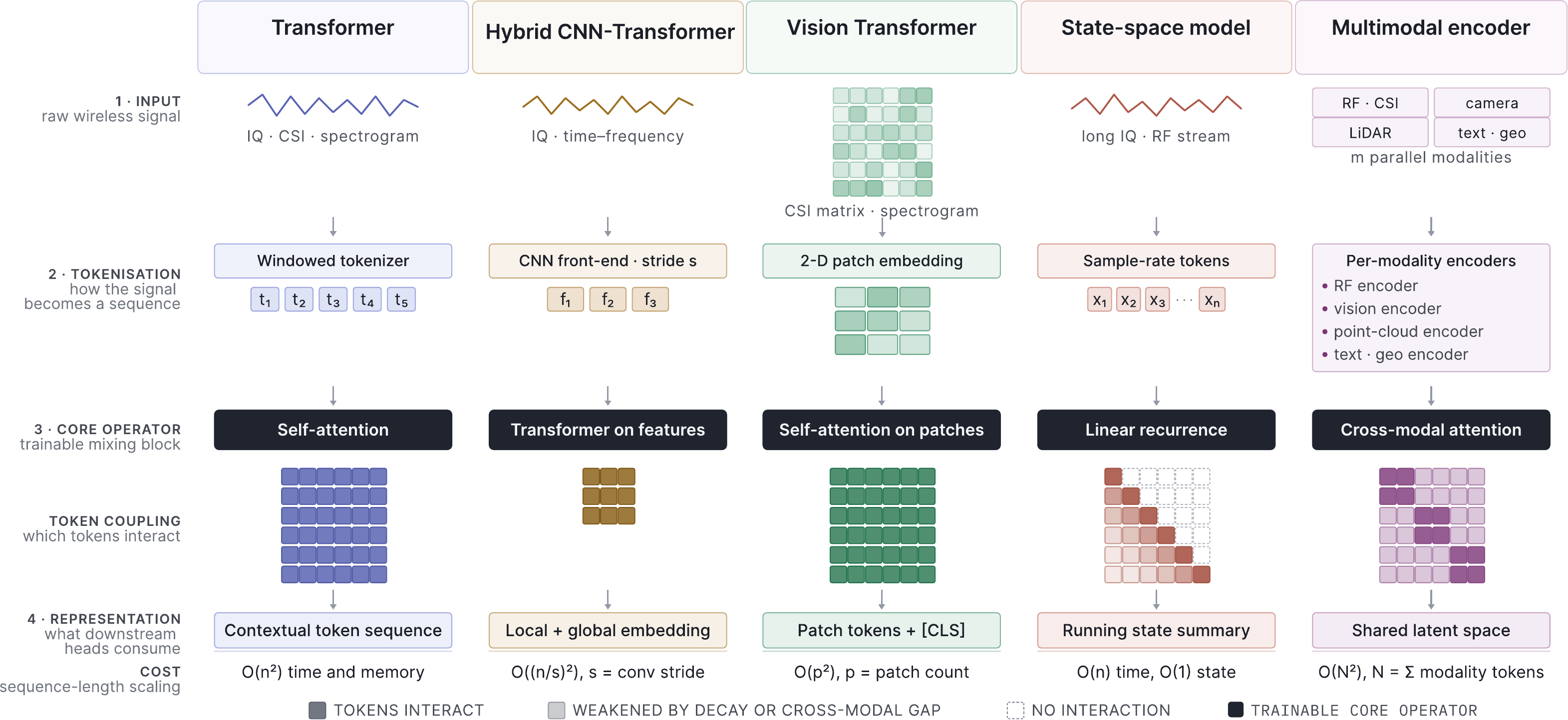}
    \caption{Neural architecture families used in WPFMs and their signal-processing pipelines: (a) Transformer tokenization and self-attention, (b) hybrid CNN--Transformer local and global processing, (c) ViT patch embedding, (d) linear-complexity state-space recurrence, and (e) multimodal encoding and fusion.}
    \label{fig:architecture_families}
\end{figure*}

The choice of architecture in a \ac{wpfm} is not only a matter of representational capacity; it is driven by wireless-specific structural requirements such as the separability of physical dimensions, deployment-time latency budgets, and the nature of the input signal. Fig.~\ref{fig:architecture_families} illustrates the main architecture families discussed below.

\textbf{Pure Transformers} are the most widely adopted backbone. LWM~\cite{alikhani2024largewirelessmodellwm} employs a 12-layer encoder-only Transformer (600K parameters) whose self-attention simultaneously captures all pairwise dependencies across antenna-subcarrier dimensions, a capability that sequential models like RNNs and spatially local models like \glspl{cnn} lack. However, the axes of a \ac{csi} matrix including time, frequency, and antenna space, represent physically distinct domains with different correlation structures. WirelessGPT~\cite{11278185} addresses this by decomposing attention into three separate heads operating along the temporal, frequency, and spatial axes independently ($\mathrm{Att}_{\mathrm{combined}} = \sum_{i \in \{T,F,S\}} \alpha_i \, \mathrm{Att}_i$), thereby preventing one dominant dimension from diluting the features of another. This decomposition is a departure from standard Transformers that process a single homogeneous text sequence.

\textbf{\ac{moe} Transformers} extend the pure Transformer with conditional computation. WiFo-E~\cite{wen2026wifoescalablewirelessfoundation} adopts sparse \ac{moe} routing to mitigate task interference during multi-task training across heterogeneous system configurations (varying antenna counts, user numbers, pilot lengths), activating only the relevant expert subset per input. WiFo-2~\cite{liu2025foundationmodelintelligentwireless} scales this to 101M parameters, and 11.6 billion \ac{csi} points while maintaining sub-linear inference cost. LWM-Spectro~\cite{kim2026lwmspectrofoundationmodelwireless} takes a protocol-specialized \ac{moe} route, training separate expert encoders for WiFi, LTE, and 5G spectrograms and routing each input to the most relevant expert, that is a design justified by the fundamentally different time-frequency signatures of each standard. On the other hand, WiFo-MiSAC~\cite{liu2026wifomisacwirelessfoundationmodel} introduces a Shared-Specific Disentangled MoE (SS-DMoE) that explicitly factorizes representations into modality-shared experts capturing cross-modal environmental semantics and modality-specific experts, with environment-context-aware routing conditioned on environment descriptors. Their analysis shows that different \ac{moe} experts activate for different carrier frequencies, demonstrating frequency-selective specialization, a phenomenon unique to wireless that has no parallel in \ac{nlp} or vision \ac{moe}.

\textbf{Hybrid \ac{cnn}-Transformer designs} exploit the complementary strengths of their two constituent modules, namely a convolutional front-end and a transformer backend, for wireless data. WirelessGPT~\cite{11278185} uses a multi-scale convolutional patch embedding module: small-scale \ac{cnn} kernels capture short-term fading and Doppler effects, while large-scale pooling captures long-term channel evolution. These multi-resolution features are then fused before entering the Transformer, addressing the fact that \ac{csi} from different propagation scenarios inherently exhibits different spatial, temporal, and frequency scales.
WirelessJEPA~\cite{chu2026wirelessjepamultiantennafoundationmodel} adopts a pure \ac{cnn} backbone (ShuffleNetV2) with sparse convolutions that operate only on unmasked patches, demonstrating that representation quality depends more on the pretraining methodology than on the model scale itself, a finding that challenges the Transformer-centric solutions.

\textbf{\ac{vit} variants} treat \ac{csi} matrices or spectrograms as images. WiFo~\cite{Liu2025} employs a \ac{vit}-based masked autoencoder with 3D convolution-based patch embedding that processes time, frequency, and space dimensions jointly. CSI-MAE~\cite{jiang2026csimaemaskedautoencoderbasedchannel} similarly uses a \ac{vit}-Base encoder with a 75\% masking ratio, far higher than the 15\% in LWM, deliberately chosen to prevent the model from taking local interpolation shortcuts and force it to learn global channel structure instead. MapViT~\cite{hsu2026mapvittwostagevitbasedframework} applies a two-stage \ac{vit} to depth maps, exploiting self-attention's ability to model long-range spatial dependencies in radio propagation maps: radio signals interact with distant objects via reflections and refractions, a phenomenon that \glspl{cnn} with local receptive fields systematically miss.
6G~WavesFM~\cite{11131142} uses a shared \ac{vit} backbone with task-specific \ac{mlp} heads across sensing, communication, and localization, demonstrating that a single \ac{vit} can serve as a multi-task physical-layer backbone when pretrained on real-world data.

\textbf{\ac{ssm}-based and hybrid attention-SSM architectures} offer linear-time complexity while preserving long-context modeling. WiMamba~\cite{raviv2026wimambalinearscalewirelessfoundation} adopts a bidirectional Mamba design in which forward and backward SSM branches with concatenated outputs, because \ac{csi} matrices lack the inherent causal ordering of language sequences; a unidirectional model would impose an undesirable causal bias on the antenna-subcarrier structure. In benchmarks, WiMamba matches or outperforms Transformer-based LWM across all four downstream tasks while achieving a 15.3$\times$ latency reduction and 42.3$\times$ memory reduction on large channel matrices. This accuracy is because of selective \ac{ssm}, whose input-dependent state transitions keeps the representation needed to model long-range antenna-subcarrier dependencies at linear cost, while its continuous-time formulation is better matched to the analog nature of \ac{rf} signals than discrete tokenized attention. Furthermore, WiMamba introduces an adaptive granularity mechanism: it supports multiple token embedding lengths $\{L_e\}$ during both training and inference, enabling deployment-time resolution-latency tradeoffs without any architectural modification. ComHymba~\cite{yang2026comhymbalowcomplexitydomaininformedfoundation} takes this further by fusing windowed self-attention with Mamba-2 \glspl{ssm} in parallel within each layer (Hymba blocks): the attention branch captures local phase fluctuations as ``snapshot memory,'' while the \ac{ssm} branch models long-range fading evolution. This hybrid yields 1.3--3.3$\times$ inference speedup over pure Transformers, growing with input dimensionality, and supports 8 downstream tasks across communication, sensing, and beam management with a single pretrained backbone.

\textbf{Spiking neural network (SNN) hybrids} introduce bio-inspired computation for noise-robust channel modeling. SpikeWFM~\cite{jing2026spikewfmspikingaidedwirelessfoundation} replaces the standard feed-forward network in each Transformer layer with a Spiking Feed-Forward Network (SFFN) using Leaky Integrate-and-Fire (LIF) neurons. The theoretical motivation is threefold: sub-threshold noise fluctuations are naturally dissipated by the leak factor (eliminating the DC bias of ReLU), coherent signal accumulates linearly over $T$ timesteps while noise grows as $\sqrt{T}$ (providing linear SNR gain), and the binary spike representation imposes an information bottleneck that prevents overfitting to environment-specific artifacts. In cross-city zero-shot channel prediction benchmarks, SpikeWFM maintains superior performance over standard Transformer baselines at moderate-to-high SNR. This performance gain stems directly from two key mechanisms. First, the coherent accumulation of spikes provides a significant advantage once a consistent signal is established. Second, the spike-induced information bottleneck prevents the model from overfitting to city-specific artifacts, thereby preserving its accuracy on unseen cities.

\textbf{Diffusion-based architectures} have emerged for tasks where the target output distribution is complex and multi-modal. WiFo-MUD~\cite{yang2026wifomudwirelessfoundationmodel} combines a Diffusion Transformer with communication-aware consistency distillation, reducing the iterative reverse diffusion process to single-step inference. This is motivated by the multi-user demodulation problem, which is NP-hard: diffusion captures the complex, multi-modal distribution of possible transmitted symbols under inter-user interference, while consistency distillation ensures the model meets real-time latency constraints.

\subsection{Input Modality and Tokenization}
\label{subsec:tax_modality}

Unlike \ac{nlp} where text is naturally discrete and images have standardized pixel grids, wireless signals present a fundamental tokenization challenge: the data is continuous, complex-valued, high-dimensional, and its structure varies with system configuration. How a \ac{wpfm} tokenizes its input directly determines what physical information the model can access and how effectively it generalizes.

\subsubsection{\ac{csi} Matrices}
\ac{csi} is the dominant input modality, adopted by LWM, WiFo, WiFo-2, CSI-MAE, WiMamba, HeterCSI, ContraWiMAE, and others. The standard approach separates the complex \ac{csi} tensor $\mathbf{H} \in \mathbb{C}^{T \times K \times N}$ (time $\times$ subcarriers $\times$ antennas) into real and imaginary components and applies non-overlapping patching. However, the specific patching strategy matters significantly for wireless performance:

\begin{itemize}
\item \textbf{1D flat patching}: LWM~\cite{alikhani2024largewirelessmodellwm} flattens the \ac{csi} matrix and divides it into patches of length 16 with learnable positional encoding, treating real and imaginary parts as separate but linked entities with coupled masking to prevent information leakage between them. This approach is simple but discards the multi-dimensional physical structure.

\item \textbf{3D patching with physics-aware positional encoding}: WiFo~\cite{Liu2025} employs 3D convolutions (Conv3d) with patch size $(t,k,n)=(4,4,4)$ along time, frequency, and space dimensions, producing variable-length token sequences from \ac{csi} of arbitrary dimensions. A Space-Time-Frequency Positional Encoding (STF-PE) assigns separate sinusoidal encodings for each physical dimension, concatenated along the feature axis. An ablation study confirms that this absolute sinusoidal encoding outperforms learnable positional embeddings for zero-shot generalization across varying \ac{csi} sizes, because learnable embeddings overfit to training-time array dimensions.

\item \textbf{Heterogeneous-scale patching}: HeterCSI~\cite{zhang2026hetercsichanneladaptiveheterogeneouscsi} confronts a harder problem that does not exist in image or text domains where input sizes are typically standardized. The \ac{csi} tensor dimensions $(T, K, A)$ themselves vary drastically across configurations (sub-6\,GHz small arrays vs.\ mmWave massive \ac{mimo}, narrowband vs.\ wideband). It uses fixed-size 3D patches producing variable-length sequences, with zero-padding to the batch-wise maximum length and double masking (attention masking to isolate valid signals from padding artifacts, combined with standard \ac{mae-model} random masking).

\end{itemize}

Real-imaginary separation (rather than magnitude-phase decomposition) is the dominant convention. ContraWiMAE~\cite{guler2025multitaskfoundationmodelwireless} explicitly justifies this: the real/imaginary representation preserves linearity and avoids phase-wrapping discontinuities at $\pm\pi$ boundaries, which would introduce artificial sharp gradients that damage both reconstruction losses and contrastive embedding quality. SPA-MAE~\cite{chen2026spamaephysicsguidedcsifoundation} offers a refinement: it applies $\mu$-law compression to the \ac{csi} magnitude before real/imaginary separation, stabilizing the dynamic range and enabling more effective reconstruction.
ComHymba~\cite{yang2026comhymbalowcomplexitydomaininformedfoundation} introduces 3D \ac{rope} that partitions the embedding dimension into three orthogonal subspaces for time, frequency, and spatial axes with decoupled rotational transformations, avoiding the fixed-resolution limitations of absolute sinusoidal encodings.
LatentWave~\cite{mohamed2026latentwavejepapretrainingwireless} proposes per-channel patch embedding: each antenna channel is independently divided into non-overlapping $16\times16$ patches through a shared linear projection. This decouples the architecture from the number of input antennas, and stochastic channel sampling during pretraining (randomly dropping antenna subsets) acts as natural data augmentation for varying array configurations.

\subsubsection{\ac{iq} Samples}
Raw complex baseband signals capture the full physical information before any frequency-domain processing, making them the most general input modality. However, \ac{iq} streams are continuous, high-rate, and lack discrete structure, making tokenization considerably harder than for \ac{csi}.

EMind~\cite{luo2025emindfoundationmodelmultitask} divides continuous \ac{iq} streams into fixed-length local patches preserving the phase-amplitude coupling and introduces a dedicated sampling-rate token that remains always unmasked during pretraining. This is a wireless-specific design: the sampling rate is an intrinsic meta-attribute spanning kHz to MHz that fundamentally governs signal interpretation, which means that a pulse width that spans 10 samples at 1\,MHz carries different physical meaning at 10\,MHz. By making the sampling rate always visible, the model naturally aligns its learned representations with rate-sensitive physical parameters such as pulse width, pulse repetition interval, and bandwidth.

WirelessJEPA~\cite{chu2026wirelessjepamultiantennafoundationmodel} processes raw multi-antenna \ac{iq} data ($\mathbf{x} \in \mathbb{R}^{2 \times 4 \times 256}$, representing \ac{iq} channels $\times$ antennas $\times$ time samples) and applies antenna-dimension upsampling—each of the 4~antenna rows is replicated 64$\times$ via nearest-neighbor interpolation to form a square antenna-time plane $\mathbf{x} \in \mathbb{R}^{2 \times 256 \times 256}$. This enables direct application of 2D block masking strategies designed for square inputs, transforming the antenna-time structure into a geometry amenable to proven image-domain masking.

\subsubsection{Spectrograms}
Time-frequency spectrograms bridge the gap between raw \ac{iq} and processed \ac{csi}. LWM-Spectro~\cite{kim2026lwmspectrofoundationmodelwireless} converts raw \ac{iq} baseband signals to log-scaled power spectrograms via \ac{stft}, then divides them into non-overlapping $4\times4$ patches. The spectrogram representation simultaneously captures modulation signatures and channel-induced effects such as Doppler broadening and multipath-induced spectral variations, making it well-suited for cross-standard signal classification across WiFi, LTE, and 5G.

\subsubsection{Multimodal Inputs}
Several models fuse \ac{rf} data with heterogeneous environmental information. ChannelGPT~\cite{11180834} combines \ac{csi} with multi-view RGB images, point clouds, and GPS positions via modality-specific embeddings fused through channel attention and environment attention mechanisms, enabling proactive channel prediction from environmental sensing. WiFo-MiSAC~\cite{liu2026wifomisacwirelessfoundationmodel} processes \ac{csi}, \ac{fmcw} radar (converted to Range-Angle and Range-Velocity maps via 2D FFT), and \ac{bev} maps simultaneously, with modality-specific linear projections, learnable modality identifiers, and modality-wise 2D positional embeddings that distinguish tokens from each modality within the shared attention.

The multimodal setting introduces the critical challenge of \textbf{temporal misalignment}, in which wireless channels vary every few milliseconds, while cameras and LiDAR operate at 10--40\,fps~\cite{zhang2026wifom2plugandplaymultimodalsensing}. WiFo-M$^2$~\cite{zhang2026wifom2plugandplaymultimodalsensing} specifically addresses this gap with a temporal sequence-to-sequence model that extrapolates future channel-aligned sensing features from historical sensor data, filling in missing environmental context between sensor frames. \cite{aboulfotouh2026multimodalwirelessfoundationmodels} introduces learned per-antenna embeddings for multi-antenna \ac{iq} tokens, arguing that standard sinusoidal positional embeddings alone cannot capture inter-antenna relationships arising from beamforming and propagation coupling.

\subsection{Pretraining Objective}
\label{subsec:tax_pretrain}

Section~\ref{subsec:ssl} introduced the general families of \ac{ssl} objectives. Here, we examine how the surveyed \glspl{wpfm} adapt these objectives to the characteristics of wireless signals, revealing design choices.

\subsubsection{Masked Modeling Adaptations}
The most used pretraining objective, adopted by LWM, WiFo, CSI-MAE, WiMamba, 6G~WavesFM, HeterCSI, EMind, and others, all use regression-based reconstruction (MSE) rather than cross-entropy prediction, since wireless signals are continuous-valued with no discrete vocabulary.

A critical but underexplored design variable is the masking ratio. LWM~\cite{alikhani2024largewirelessmodellwm} uses only 15\% masking (mirroring BERT), whereas CSI-MAE~\cite{jiang2026csimaemaskedautoencoderbasedchannel} and 6G~WavesFM~\cite{11131142} employ 75\%. CSI-MAE explicitly justifies the high ratio: wireless channels exhibit strong spatial correlations across adjacent subcarriers and antennas, so a low masking ratio allows the model to reconstruct masked patches via trivial local interpolation rather than learning deep physical structure. The 75\% ratio forces reconstruction from distant, structurally informative tokens, yielding representations that capture global propagation patterns.

WiFo~\cite{Liu2025} introduces a triple-objective masking strategy that goes beyond isotropic random masking:
(1)~Random masking at a 85\% ratio captures 3D structured features across the full space-time-frequency volume;
(2)~Time masking at a 50\% ratio masks future tokens to learn temporal channel evolution; and
(3)~Frequency masking at a 50\% ratio masks frequency tokens to learn adjacent-band variations.
Ablation studies demonstrate that all three are essential: removing time masking improves frequency prediction but destroys time prediction (and vice versa), whereas removing random masking degrades both. This multi-objective masking strategy is motivated by the multi-dimensional structure of wireless channels, where temporal, spectral, and spatial correlations follow distinct physical laws (Doppler dynamics, frequency-selective fading, and array steering, respectively).

WiFo-MiSAC~\cite{liu2026wifomisacwirelessfoundationmodel} further introduces \ac{csi}-specific masking patterns: random masking for robustness to arbitrary missing data, frequency-domain masking to recover structured spectral correlations, and comb masking that mimics the comb-type pilot sampling patterns used in OFDM systems, directly bridging self-supervised pretraining with practical wireless protocol constraints.

EMind~\cite{luo2025emindfoundationmodelmultitask} addresses a distinct challenge: signals from heterogeneous electromagnetic sources (communication, radar, interference) span 32$\times$ variation in length. Rather than zero-padding to a uniform length, which wastes computation, EMind employs multi-signal packing, concatenating multiple variable-length \ac{iq} samples into ultra-long sequences with per-sample masking and boundary indices to prevent cross-sample interference. This hardware-aware strategy eliminates the massive padding overhead of traditional batching.

ComHymba~\cite{yang2026comhymbalowcomplexitydomaininformedfoundation} introduces domain-specific masking strategies that go beyond random or dimension-specific patterns: (a)~multi-scale random masking, (b)~dimension-specific masking along time, frequency, or antenna axes, (c)~pilot-pattern masking that mimics actual reference signal layouts used in standards, and (d)~deep-fading block masking that simulates path blockage events. A progressive mask ratio curriculum (50\%$\to$75\%) is applied alongside a domain-informed loss that decouples amplitude consistency and phase alignment from the standard statistical reconstruction, with physical losses activated after the model reaches a statistical plateau.

SPA-MAE~\cite{chen2026spamaephysicsguidedcsifoundation} proposes physics-guided pretraining that augments standard masked reconstruction with two auxiliary branches: a structure-aware branch that predicts the 2D FFT magnitude of the CSI (exploiting the angular-delay domain sparsity via a compressed-sensing-inspired soft-thresholding head), and a parameter-aware branch that aligns a learnable SPA token with contrastively pretrained multipath parameter representations (delay, path power, angles). Stage-wise training prevents interference between reconstruction and physics-guided objectives. Remarkably, this physics-guided approach achieves better downstream performance with fewer parameters.

The surveyed masked modeling approaches differ along three principal design axes of masking ratio, masking geometry, and loss formulation, each of which exposes a distinct trade-off:

\begin{itemize}
\item \textbf{Masking ratio determines representation depth vs.\ reconstruction stability.} Low ratios ($\sim$15\%, LWM) preserve most of the input context, facilitating stable training and accurate patch-level reconstruction, but risk learning shallow representations. High ratios (75--85\%, CSI-MAE, WiFo, 6G~WavesFM) force the model to infer global propagation structure from distant tokens, yielding richer representations at the cost of a more challenging optimization landscape.

\item \textbf{Masking geometry controls inductive bias.} Isotropic random masking (LWM, CSI-MAE, WiMamba) makes no assumptions about which physical dimension matters most, producing general-purpose representations. In contrast, dimension-specific masking introduces inductive biases aligned with specific wireless phenomena. Multi-objective masking is necessary for models intended to serve diverse downstream tasks, whereas a single focused masking geometry may be preferable when the target task family is known in advance.

\item \textbf{Input heterogeneity requires architectural innovations beyond masking design.} EMind's multi-signal packing addresses the practical challenge of highly variable signal lengths across electromagnetic sources. HeterCSI's double-masking strategy similarly tackles the heterogeneity of \ac{csi} tensor dimensions across system configurations. These solutions highlight that effective masked pretraining for wireless foundation models requires co-designing the masking strategy.
\end{itemize}

\subsubsection{Contrastive Objectives}
ContraWiMAE~\cite{guler2025multitaskfoundationmodelwireless} demonstrates that combining contrastive and reconstructive objectives is better for wireless channels. Its key innovation is to use noise as natural augmentation: positive pairs are created by adding AWGN to the same channel and applying different random masks to the original and noisy versions. This exploits the inherent noise and partial observability of wireless signals as natural augmentation, eliminating the need for synthetic augmentation. The results are invariant to specific antenna/subcarrier selections and noise conditions, properties directly relevant to practical deployments where channel observations are always noisy and partial. In benchmarks, ContraWiMAE achieves top-1/top-3 beam selection accuracy of 72.1\%/87.2\% with only 10\% training data, compared to 25.3\%/36.1\% for LWM under identical conditions.

\subsubsection{Predictive and Generative Objectives}
WirelessJEPA~\cite{chu2026wirelessjepamultiantennafoundationmodel} departs from both masked reconstruction and contrastive learning by predicting masked latent representations rather than raw signal values. This \ac{jepa} approach~\cite{assran2023self}, which was introduced for images and predicts the latent representations of masked regions from the visible context in an abstract embedding space, avoids pixel-level reconstruction artifacts and focuses on learning semantic/structural features. A key finding from WirelessJEPA is that masking geometry directly controls inductive bias: time masks favor waveform-level tasks such as modulation classification, antenna masks favor spatial tasks such as angle-of-arrival estimation, and multi-block masks provide the best compromise. This is uniquely wireless: the choice of which physical dimension to mask creates a tunable bias aligned with the patterns of the target task group.

LatentWave~\cite{mohamed2026latentwavejepapretrainingwireless} extends the \ac{jepa} paradigm to support both spectrograms and CSI in a unified framework, confirming the masking-geometry finding: frequency masking strongly favors channel-related tasks but hurts RF signal classification (a 15 percentage point drop), establishing that no single masking strategy dominates across all wireless tasks. CSI-JEPA~\cite{luo2026csijepafoundationrepresentationsubiquitous} introduces channel variation-aware masking for Wi-Fi sensing: rather than uniform sampling, it estimates temporal and subcarrier-domain signal variations to guide target region selection toward the most informative parts of the CSI, achieving up to 98\% label savings compared to fully supervised baselines across 7 diverse sensing tasks.

\subsection{Model Scale and Deployment Target}
\label{subsec:tax_scale}

An important characteristic of \glspl{wpfm} compared to foundation models in \ac{nlp} or vision is that ``large'' refers primarily to capability rather than parameter count~\cite{cheng2026largewirelessfoundationmodels}. Most existing \glspl{wpfm} operate with fewer than 100M parameters; WiFo-2 at 101M~\cite{liu2025foundationmodelintelligentwireless} and WirelessGPT at 80M~\cite{11278185} represent the upper end, while LWM achieves competitive results with only 600K parameters~\cite{alikhani2024largewirelessmodellwm}. This is in contrast to \ac{nlp} foundation models that routinely exceed 10 billion parameters. We hypothesize two reasons: wireless channel data is high-dimensional and exhibits stronger physical structure (governed by Maxwell's equations and propagation physics) than natural language, meaning smaller models can capture the relevant dynamics; and the deployment targets (sub-millisecond latency and edge compatibility) impose constraints that large models cannot satisfy.

The mismatch between model capacity and wireless hardware was quantified in Section~\ref{subsec:sizing_guidelines}. A typical \ac{ue} baseband ASIC processing channel estimation within a fraction of an OFDM symbol duration possesses approximately 2\,MB of SRAM, restricting deployable models to roughly 1.7 million INT8 parameters. This makes aggressive compression and knowledge distillation an operational necessity for edge deployment.

\cite{cheng2026largewirelessfoundationmodels} identifies two deployment paradigms: (1)~leveraging general-purpose foundation models by cross-modal transfer (e.g., fine-tuning GPT-2 for channel prediction, as in ChannelGPT~\cite{11180834}), suitable for latency-tolerant tasks at the base station; and (2)~building wireless-native foundation models with purpose-built architectures (e.g., WiFo, LWM, WiMamba), which support more tasks with fewer parameters and are suitable for latency-sensitive inner-loop processing.

WiMamba~\cite{raviv2026wimambalinearscalewirelessfoundation} illustrates the importance of architectural efficiency: at 2.5M parameters, it achieves comparable accuracy to Transformer-based LWM while reducing inference latency from 363\,ms to 23\,ms and memory consumption from 4.8\,GB to 114\,MB on large channel matrices. Initial scaling studies provide mixed evidence. \cite{buffelli2025towards} report improved performance when model size (5M$\to$100M), dataset size (1M$\to$10M), and training compute increase together. In contrast, CSI-MAE~\cite{jiang2026csimaemaskedautoencoderbasedchannel} finds that a larger model degrades by 109\% in antenna-domain extrapolation while improving by 54.8\% in subcarrier-domain extrapolation. HeterCSI further reports improved zero-shot performance as the number of heterogeneous pretraining datasets increases, provided that scale conflicts are controlled~\cite{zhang2026hetercsichanneladaptiveheterogeneouscsi}. These results suggest that data composition and the physical transfer axis matter as much as parameter count.

\cite{cheng2026bigwirelessfoundationmodel} evaluate MAE architectures from 104K to 150M parameters across 10 channel datasets and relate saturation to the estimated intrinsic dimensionality of the channel data. In their setting, scaling from 12M to 96M parameters yields only $\sim$2.3\,dB improvement, whereas a 12M model with five steps of pilot-aided \ac{ttt} achieves a 7.2\,dB improvement. This result supports right-sizing and adaptation for channel estimation, but it does not establish a universal scaling law across architectures, modalities, and downstream tasks. Wireless scaling laws therefore remain an open empirical question, as discussed in Section~\ref{subsec:future_scaling}.

\subsection{Generalization Capability}
\label{subsec:tax_generalization}

We classify generalization along three axes: \textbf{cross-device} (different antenna hardware), \textbf{cross-band} (different frequency ranges), and \textbf{cross-region} (different propagation environments). Models are evaluated on their ability to transfer without additional fine-tuning (zero-shot) or with minimal adaptation (few-shot). 

\subsubsection{Evidence for Zero-Shot Generalization}

In this survey, zero-shot generalization means that all weights required for the evaluated task, including the backbone and task head, remain fixed and that no labeled or unlabeled samples from the target domain are used for optimization. The target may introduce an unseen site, frequency band, antenna configuration, channel scale, or modality combination. This setting is different from task-level zero-shot transfer: most reported wireless experiments retain a known task and test it under an unseen domain. Evidence for performing a completely new task without training a task-specific head remains limited. Few-shot evaluation, in contrast, permits a small target-domain set for adaptation.

For cross-configuration transfer, WiFo~\cite{Liu2025} is trained on datasets spanning multiple NR bands, propagation scenarios, and mobility ranges, and is then evaluated without target-domain retraining. WiFo-2 extends this protocol to channel reconstruction under unseen configurations and \gls{csi} distributions~\cite{liu2025foundationmodelintelligentwireless}. Beyond the WiFo family, HeterCSI evaluates 12 held-out datasets without scenario-specific fine-tuning and shows that scale-aware heterogeneous pretraining improves over the WiFo zero-shot baseline~\cite{zhang2026hetercsichanneladaptiveheterogeneouscsi}. These studies indicate that both pretraining diversity and batch construction affect transfer.

Other models provide evidence along different axes. CSI-MAE~\cite{jiang2026csimaemaskedautoencoderbasedchannel} demonstrates cross-frequency transfer from RMa at 2.4\,GHz to 0.7 and 3.5\,GHz without target-frequency adaptation. SpikeWFM~\cite{jing2026spikewfmspikingaidedwirelessfoundation} evaluates cross-city channel prediction. WiFo-MiSAC~\cite{liu2026wifomisacwirelessfoundationmodel} applies a multimodal model to unseen links without fine-tuning; separate experiments study robustness to missing modalities and the integration of new modalities. WirelessJEPA~\cite{chu2026wirelessjepamultiantennafoundationmodel} reports out-of-distribution transfer with a frozen encoder, although its linear-probing protocol uses downstream labels and is therefore better classified as few-shot representation transfer rather than strict zero-shot inference. Similarly, retrieval-based localization that uses a target-scene fingerprint database is training-free but not zero-shot under the strict definition above. These protocol differences should be stated explicitly when comparing models.

\subsubsection{The Gradient Dynamics of Heterogeneous Training}
HeterCSI~\cite{zhang2026hetercsichanneladaptiveheterogeneouscsi} provides a fundamental insight into why some forms of data diversity help generalization while others hinder it. Through gradient analysis, they demonstrate that when \ac{csi} samples with different scales (e.g., 4-antenna narrowband and 64-antenna wideband) are batched together during training, 85.96\% of gradient cosine similarities are negative, indicating destructive gradient interference. In contrast, when samples share the same scale but come from different propagation scenarios, fewer than 0.10\% exhibit negative similarity, indicating constructive gradient alignment. This finding has a direct design implication: scenario diversity during pretraining is beneficial, while scale heterogeneity must be carefully managed through adaptive batching strategies that group same-scale samples together. HeterCSI's sort-and-partition heuristic achieves optimal minimum padding overhead and reduces training latency by 53\% compared to global shuffling.

\subsubsection{Prompting a Wireless Foundation Model}
A fundamental barrier to zero-shot generalization is the absence of a discrete, natural vocabulary. Unlike \ac{nlp}, where models are explicitly guided by text prompts, ``prompting'' a wireless foundation model requires innovative techniques. WiFo's STF-PE encodes the physical dimensions of the target system as positional information, enabling the model to adapt to unseen \ac{csi} configurations. ChannelGPT~\cite{11180834} injects environmental context (GPS, images, point clouds) as side information, demonstrating that only fine-tuning LayerNorm and positional encoding parameters suffices for cross-scenario, cross-frequency sim-to-real adaptation, achieving a 49.67\% \ac{rmse} improvement over \glspl{dnn} with only 10\% real-world samples. 6G~WavesFM~\cite{11131142} confirms that domain-aligned pretraining is critical: misaligned pretraining data hurts downstream performance (a model pretrained on 5G \ac{csi} performs poorly on non-OFDM tasks), demonstrating negative knowledge transfer, a risk largely absent in \ac{nlp} where diverse texts improve performance.

\subsection{Overview and Key Takeaways}
\label{subsec:tax_synthesis}

The taxonomy reveals several recurring themes that distinguish \glspl{wpfm} from foundation models in other domains:

\begin{itemize}
\item \textbf{Physics-aware design tends to help.} Many of the stronger models reviewed here embed wireless domain structure into their architecture and training procedure, for example triple-axis attention (WirelessGPT~\cite{11278185}), physics-aware positional encoding (WiFo~\cite{Liu2025}), bidirectional processing for non-causal \ac{csi} (WiMamba~\cite{raviv2026wimambalinearscalewirelessfoundation}), task-aligned masking geometries (WirelessJEPA~\cite{chu2026wirelessjepamultiantennafoundationmodel}), and protocol-aware comb masking (WiFo-MiSAC~\cite{liu2026wifomisacwirelessfoundationmodel}). Several of these studies report that directly transferring \ac{nlp} or vision architectures without such adaptations gives weaker results~\cite{11131142}, although a systematic head-to-head comparison across models is still lacking.

\item \textbf{Capability may scale differently than in \ac{nlp}.} The surveyed \glspl{wpfm} achieve competitive results at 0.6--100M parameters, and the reported scaling behavior appears dimension-dependent rather than uniformly monotonic~\cite{buffelli2025towards, cheng2026bigwirelessfoundationmodel}. In particular, CSI-MAE~\cite{jiang2026csimaemaskedautoencoderbasedchannel} observed that larger models can degrade on some extrapolation axes (e.g., the antenna domain) while improving on others. These observations come from a small number of studies and are best read as emerging evidence rather than established scaling laws.

\item \textbf{Data diversity may need careful management.} Gradient analysis in HeterCSI~\cite{zhang2026hetercsichanneladaptiveheterogeneouscsi} indicates that scenario diversity can be constructive while scale heterogeneity can cause destructive gradient interference, motivating wireless-specific data strategies (adaptive batching, multi-signal packing) beyond standard data augmentation. This evidence is so far specific to \ac{csi} pretraining and has not been broadly validated across modalities.

\item \textbf{A deployment division is starting to emerge.} Across the surveyed works, a rough two-tier pattern can be observed: higher-capacity designs such as \ac{moe} Transformers~\cite{shazeer2017outrageously}\footnote{A Mixture-of-Experts layer replaces a single dense sub-network with many parallel expert sub-networks and a gating function that routes each input to only a few of them, increasing model capacity while keeping the per-input computation roughly constant.}, diffusion models, and multimodal architectures are typically targeted at the network core and high-compute edge servers for latency-tolerant tasks, whereas \glspl{ssm}, lightweight hybrids, and distilled models are more often proposed for resource-constrained edge deployments under tight SWaP (Size, Weight, and Power) constraints. This division is a qualitative trend across current papers rather than a firmly established consensus.
\end{itemize}

\begin{table*}[htbp]
\scriptsize
\centering
\caption{Design Taxonomy of Representative WPFMs by Architecture, Scale, Input Modality, Pretraining Objective, Downstream Tasks, and Generalization}
\label{tab:comparison}
\setlength{\tabcolsep}{1pt}
\renewcommand{\arraystretch}{1.3}
\begin{tabular*}{\textwidth}{@{\extracolsep{\fill}}>
{\bfseries\raggedright\arraybackslash}m{1.4cm} >{\raggedright\arraybackslash}m{0.55cm} >{\raggedright\arraybackslash}m{2.0cm} >{\raggedright\arraybackslash}m{0.75cm} >{\raggedright\arraybackslash}m{1.55cm} >{\raggedright\arraybackslash}m{1.6cm} >{\raggedright\arraybackslash}m{3.1cm} >{\raggedright\arraybackslash}m{3.25cm} >{\raggedright\arraybackslash}m{1.65cm} @{}}
\toprule
\textbf{Model} & \textbf{Ref.} & \textbf{Architecture} & \textbf{Params}
    & \textbf{Input Modality} & \textbf{Pretraining Obj.}
    & \textbf{Reported Data} & \textbf{Downstream Tasks}
    & \textbf{Generalization} \\
\midrule
LWM           & \cite{alikhani2024largewirelessmodellwm}         & Transformer        & 0.6M    & CSI             & Masked            & $>$1M channels, 15 DeepMIMO scenarios & Beam prediction, LoS/NLoS  & Cross-scenario   \\
\rowcolor[gray]{0.94}
WiFo          & \cite{Liu2025}             & ViT    & 0.3--86M    & CSI             & Masked            & 160K samples, 16 QuaDRiGa configurations & Channel prediction           & Cross-config    \\
WiFo-2        & \cite{liu2025foundationmodelintelligentwireless} & MoE Transformer    & 101M  & CSI             & Masked + Denoising & 11.6B CSI points, 78 sub-datasets & Channel pred., est., feedback, localization & Zero-shot + few-shot\\
\rowcolor[gray]{0.94}
WiFo-E        & \cite{wen2026wifoescalablewirelessfoundation}            & MoE Transformer    & 11M    & CSI             & Multi-task        & 300K channels, 9 QuaDRiGa configurations & FDD precoding                & Cross-config    \\
WiFo-M2       & \cite{zhang2026wifom2plugandplaymultimodalsensing}         & Multimodal         & 0.07--9.6M    & LiDAR+Image  & Contrastive        & M3SC/SynthSoM multimodal data & Sensing+Comm                 & Cross-scenario   \\
\rowcolor[gray]{0.94}
WiFo-MUD      & \cite{yang2026wifomudwirelessfoundationmodel}         & Diffusion Transformer & 1.3--5.8M    & IQ+CSI              & Diffusion            & Multi-user       & Demodulation                 & Cross-config    \\
WiFo-CF       & \cite{11479626}         & S-R MoE Transformer   & 0.1--14M     & CSI (FDD)           & Masked (self-sup.)   & LH-CDF, $>$1M samples & CSI feedback (FDD), localization & Cross-config    \\

\rowcolor[gray]{0.94}
WirelessGPT   & \cite{11278185}     & CNN+Transformer        & 80M    & CSI      & Masked (generative)        & Simulated and measured CSI & ISAC, sensing, comm          & Cross-task \\
ChannelGPT    & \cite{11180834}        & \gls{llm}-based (GPT-2)  & N/R & CSI+Context     & Generative        & Measured and ray-traced multimodal data & Channel modeling             & Cross-env    \\
\rowcolor[gray]{0.94}
LWM-Spectro   & \cite{kim2026lwmspectrofoundationmodelwireless}       & MoE (ViT)        & N/R    & Spectrograms    & Masked + Contrastive & 9.2M samples       & Signal recognition           & Cross-band    \\
CSI-MAE       & \cite{jiang2026csimaemaskedautoencoderbasedchannel}         &  ViT & N/R    & CSI             & Masked            & Sionna/3GPP channel data & Channel extrap., feedback, positioning & Cross-freq   \\
\rowcolor[gray]{0.94}
WirelessJEPA  & \cite{chu2026wirelessjepamultiantennafoundationmodel}     & CNN-JEPA                & N/R    & Multi-antenna IQ   & JEPA (predictive)  & Multi-site measured IQ & AoA, classification, fingerprinting        & Cross-task  \\
MapViT        & \cite{hsu2026mapvittwostagevitbasedframework}           & Two-stage ViT      & 0.6M    & Depth maps      & SSL + Supervised        & 1K ray-traced scenes & Radio map prediction         & Cross-env      \\
\rowcolor[gray]{0.94}
6G WavesFM    & \cite{11131142}       & ViT + MLP heads    & 38M enc.    & Spectro.+CSI+IQ & Masked            & 4,648 multi-source pretraining samples & Positioning, ch.\ est., sensing, classif.  & Cross-task    \\
EMind         & \cite{luo2025emindfoundationmodelmultitask}            & Transformer     & 110M    & EM signals (IQ)      & Masked & 81M samples     & SIGINT, classification       & Cross-domain  \\
\rowcolor[gray]{0.94}
MobiFM        & \cite{11298194}          & Diffusion+MoE Transf. & 13--88M    & Mobile data     & Generative        & 10 real-world mobile datasets & Traffic forecasting          & Cross-type     \\
HeterCSI      & \cite{zhang2026hetercsichanneladaptiveheterogeneouscsi}       & Transformer        & N/R    & Heterogeneous CSI & Masked     & 40 training + 12 held-out datasets & Channel reconst., prediction       & Cross-scale  \\
\rowcolor[gray]{0.94}
WiMamba       & \cite{raviv2026wimambalinearscalewirelessfoundation} & SSM (Mamba)        & 2.5M    & CSI             & Masked            & Multi-resolution DeepMIMO & Beam pred., LoS, localization & Cross-task      \\
FM-WTR & \cite{11264506} & Transformer & N/R & IQ+CIR & Masked & Multi-tech & Tech recognition, localization & Cross-device \\
\rowcolor[gray]{0.94}
ContraWiMAE          & \cite{guler2025multitaskfoundationmodelwireless}        & ViT        & 0.6M    & CSI             & Contrastive + MAE   & 2.5M samples       & Beam selection, LoS, ch.\ est. & Cross-scenario      \\
WiFo-MiSAC   & \cite{liu2026wifomisacwirelessfoundationmodel}  & SS-DMoE Transformer & N/R  & CSI+Radar+Map   & Masked + Contrastive & $>$1B CSI entries, $>$200K aligned triplets & Beam pred., ch.\ est., positioning & Zero-shot + few-shot \\
\rowcolor[gray]{0.94}
SPA-MAE      & \cite{chen2026spamaephysicsguidedcsifoundation}  & ViT + SPA token & 0.33M & CSI & Masked + Physics-guided & Seen + 6 unseen DeepMIMO cities & LoS/NLoS, beam pred., ch.\ est., positioning & Cross-scenario \\
ComHymba     & \cite{yang2026comhymbalowcomplexitydomaininformedfoundation} & Hymba (Att.+SSM) & 100M & CSI & Masked (domain-informed) & 198.5\,GB, 33 scenarios & 8 tasks: ch.\ pred./est., localization, beam & Cross-band \\
\rowcolor[gray]{0.94}
SpikeWFM     & \cite{jing2026spikewfmspikingaidedwirelessfoundation} & SNN-Transformer & 3.5--80M & CSI & Masked & DeepMIMO & Channel prediction & Cross-city \\
LatentWave   & \cite{mohamed2026latentwavejepapretrainingwireless} & ViT-JEPA & 6.4M & Spectro.+CSI & JEPA (latent) & Multi-modal & RF classif., positioning, beam pred. & Cross-modality \\
\rowcolor[gray]{0.94}
CSI-JEPA     & \cite{luo2026csijepafoundationrepresentationsubiquitous} & ViT-JEPA & $\sim$3M & WiFi CSI ampl. & JEPA (latent) & 151K CSI windows, 7 datasets & 7 sensing tasks (HAR, fall, breathing, etc.) & Cross-task \\
SiFo         & \cite{zhao2026sifowirelessfoundationmodel} & MLP + Memory & N/R & RSRP fingerprints & Capture-efficiency & 10 DeepMIMO cities & FDD CSI feedback & Cross-site \\
\rowcolor[gray]{0.94}
RA-LWLM      & \cite{pan2026ralwlmretrievalaugmentedincontextlocalization} & Transformer + MoE-ICL & N/R & CSI & DTI (domain-transf.) & 20 seen + 10 unseen scenes & Localization & Cross-environment \\
\bottomrule
\end{tabular*}
\vspace{1mm}

\raggedright\footnotesize{``N/R'' in the Params column indicates that the corresponding paper does not report the number of parameters. Data quantities follow the reporting units used by the source studies (samples, channel tensors, CSI points, or scenarios) and are therefore not interchangeable. The generalization column summarizes the terminology used by each study. Zero-shot protocols differ in whether the task, task head, domain, or modality is unseen; Section~\ref{subsec:tax_generalization} defines the interpretation used in this survey.}
\end{table*}

\section{Telecommunications Applications}
\label{sec:telecom}

\begin{figure*}
    \centering
    \includegraphics[width=\linewidth]{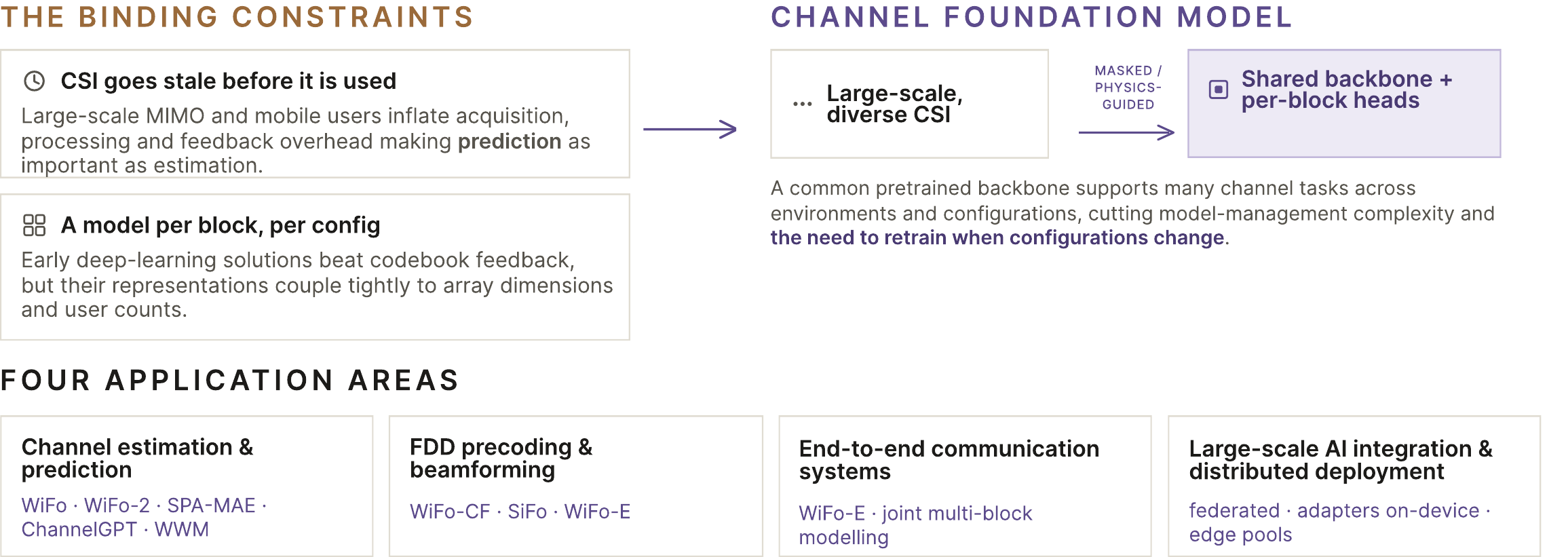}
        \caption{Foundation-model approach to telecommunication applications.}

    \label{fig:telecomm_app}
\end{figure*}
 
Telecommunications represents the most mature application domain for \glspl{wpfm}, encompassing channel modeling, beamforming, and end-to-end communication design.

\subsection{Channel Estimation and Prediction}
\label{subsec:channel}

Channel estimation and prediction are fundamental wireless communication tasks, as accurate \gls{csi} is required for equalization and beamforming. As wireless networks evolve from 5G toward 6G, the adoption of large-scale \gls{mimo} systems and support for highly mobile users significantly increase \gls{csi} acquisition, processing, and feedback overhead. In such scenarios, the estimated \gls{csi} may become outdated before it can be effectively utilized, leading to degraded transmission performance. Consequently, channel prediction is becoming as important as channel estimation. By forecasting future channel conditions, channel prediction enables proactive adaptive modulation and coding, predictive beamforming and handover management and intelligent resource allocation. These capabilities are expected to play a critical role in emerging 6G applications such as autonomous driving and \ac{xr}, where communication systems must adapt to channel variations in real time.

The growing importance of predictive wireless intelligence is also reflected in ongoing 3GPP standardization activities for 5G-Advanced and beyond. Recent Release-18 and Release-19 studies have introduced AI/ML-assisted air-interface functionalities, including \gls{csi} feedback enhancement, beam management, and positioning~\cite{lin2023overview3gppstudyartificial}~\cite{9970357}~\cite{10123939}. These activities share a common objective of exploiting temporal and contextual wireless information to improve network performance while reducing signaling overhead.

\glspl{cfm} have recently emerged as a solution for these challenges. Learning transferable wireless representations from large-scale datasets, \gls{cfm} can support multiple downstream tasks across different environments using a pretrained backbone and then applying fine-tuning. This shared backbone reduces model-management complexity by enabling multiple downstream tasks to share a single pretrained model rather than requiring multiple task-specific models and improves scalability across heterogeneous deployments by adapting the same backbone to different propagation environments and system configurations. This design is consistent with recent AI-native 6G frameworks, which support replacing multiple task-specific models with a common backbone that can be adapted to multiple downstream tasks \cite{wu2026resilientautonomousnetworksbluesky}.

Among the earliest \gls{cfm} developed for channel intelligence, WiFo~\cite{Liu2025} formulates channel prediction as a sequence modeling problem over \gls{ofdm} subcarriers and time slots. Using a masked autoencoder with a \gls{vit} backbone, WiFo demonstrates strong zero-shot generalization across diverse space-time-frequency configurations, including different system configurations and frequency bands. Building on this, \cite{liu2025foundationmodelintelligentwireless} introduce WiFo-2, a sparse-\gls{moe} \gls{cfm} which achieves zero-shot performance on channel estimation and channel prediction tasks while demonstrating strong few-shot adaptation across a broad range of downstream applications, including channel estimation and prediction. Furthermore, a hardware prototype validates its practical deployment feasibility, suggesting that \glspl{cfm} may serve as a common \gls{ai} backbone for future wireless systems.

WiFo and WiFo-2 demonstrate data-driven representation learning for channel estimation and prediction. However, purely data-driven pretraining may not fully exploit the physical principles of wireless propagation. To bridge this gap, SPA-MAE~\cite{chen2026spamaephysicsguidedcsifoundation} introduces a physics-guided \gls{cfm} that incorporates channel propagation priors directly into the pretraining process through a physical prior module. Specifically, SPA-MAE exploits both explicit multipath channel parameters and sparse transformed-domain channel structures as complementary guidance signals. The model learns meaningful representations that better reflect wireless propagation mechanisms. Experimental results demonstrate consistent gains across multiple downstream wireless tasks such as channel estimation, beam prediction, and user positioning, particularly under low \gls{snr} and limited-data conditions, illustrating the benefits of integrating wireless physics into \gls{cfm} pretraining.

Despite these advances, most existing \glspl{cfm} rely on synthetic data and focus on learning representations from channel-related information. Although models such as SPA-MAE incorporate physics-based propagation priors, they generally do not exploit environmental context. To address these limitations, recent research has extended \gls{cfm} toward environment-aware channel modeling. ChannelGPT~\cite{11180834} incorporates environmental context, including building maps and positioning information, into a \gls{llm}-style architecture for real-world 6G channel modeling. By jointly processing wireless measurements and environmental information, it demonstrates the potential of multimodal wireless intelligence. Similarly, \cite{chen2026wirelessworldmodelainative} proposes the \gls{wwm}, a physics-aware multimodal framework that models the causal relationship between environment geometry and signal propagation. \gls{wwm} jointly processes \gls{csi} matrices, 3D point clouds, and user trajectories through a \gls{jepa} equipped with a \gls{moe} Transformer. \gls{wwm} outperforms both unimodal foundation models and task-specific baselines in seen and unseen environments and in real-world measurement scenarios in different downstream tasks. Nevertheless, such multimodal \glspl{cfm} require large-scale datasets that combine wireless measurements with environmental information, which are expensive to collect. Moreover, their computational complexity may limit real-time deployment in practical wireless systems.

From an industrial and standardization perspective, ChannelGPT and \gls{wwm} represent an important step toward the realization of \gls{isac}~\cite{kaushik2023integratedsensingcommunications6g} ~\cite{10330577} and \gls{eic}~\cite{ZHANG2026186}, two key technological directions envisioned for future 6G networks. \gls{isac} enables wireless systems to extract information about the surrounding environment, such as the location and movement of objects, from radio signals to support the data transmission. \gls{eic} further incorporates environmental and contextual information, such as user location, obstacles, and mobility into communication-related decisions. Such information can be used to anticipate channel variations and adapt functions such as beam management, handover, and resource allocation, thereby improving communication reliability and efficiency. Multimodal \gls{cfm} learn unified representations of the interaction between physical environments and wireless propagation, offering a foundation for environment-aware radio systems that jointly support communication and sensing.

Despite recent advances, several challenges remain for applying \glspl{cfm} to channel estimation and prediction:
\begin{enumerate}
    \item \textbf{Limited modality coverage.} Existing multimodal \glspl{cfm} incorporate only a few modalities. Future models should integrate complementary information from LiDAR, cameras and video, optical and acoustic sensing, maps, weather, and traffic data to improve prediction accuracy. Adding more modalities, however, raises the problem of asynchronous multimodal learning: different modalities are acquired at different rates, and designing \glspl{cfm} that fuse and reason over such multi-rate observations remains open. 
    \item \textbf{Lack of continual adaptation.} Most current models cannot adapt online and degrade as environments evolve through weather variations, user mobility, and infrastructure changes. Developing online learning mechanisms is therefore an important research direction.
    \item \textbf{Scarcity of real-world data.} Progress is constrained by the lack of large-scale publicly available real-world wireless datasets, which are essential for model generalization and practical deployment.
 \end{enumerate}
\subsection{FDD Precoding and Beamforming}
\label{subsec:precoding}

Massive \gls{mimo} has significantly enhanced the performance of wireless networks by providing substantial gains in spectral efficiency and system capacity. To fully exploit these gains, the \gls{bs} requires accurate \gls{csi} to perform downlink transmission optimization~\cite{8395053}. In particular, \gls{csi} is a fundamental input for precoding and beamforming, which spatially direct transmitted signals toward intended users while mitigating inter-user interference. Consequently, the performance of \gls{mimo} systems strongly depends on the accuracy of available \gls{csi} at the transmitter.

Acquiring such \gls{csi}, however, remains a major challenge, particularly in \gls{fdd} systems. Unlike TDD operation, where channel reciprocity allows the \gls{bs} to infer downlink \gls{csi} from uplink measurements, \gls{fdd} systems require explicit \gls{csi} acquisition and feedback. Specifically, the \gls{ue} must estimate the downlink channel, compress the \gls{csi}, and feed it back to the \gls{bs}, which subsequently reconstructs the \gls{csi} and computes the corresponding precoding vectors. As antenna arrays continue to grow in size in 5G-Advanced and future 6G systems, the dimensionality of \gls{csi} increases, causing \gls{csi} feedback to consume substantial uplink resources and making \gls{csi} acquisition one of the primary bottlenecks for efficient \gls{fdd} beamforming and precoding~\cite{nerini2023machinelearningbasedcsifeedback},~\cite{9931713}.

This challenge is reflected in ongoing 3GPP standardization activities. In particular, the substantial uplink signaling overhead associated with explicit \gls{csi} acquisition and feedback in \gls{fdd} systems has motivated AI/ML-assisted \gls{csi} feedback enhancement as a representative use case for the AI-native air interface study starting from 3GPP Release 18 ~\cite{lin2023overview3gppstudyartificial},~\cite{9970357}, with the objective of reducing feedback overhead while maintaining high precoding and beamforming performance in massive \gls{mimo} deployments.

Early deep learning approaches have demonstrated significant improvements over conventional codebook-based \gls{csi} feedback methods by learning compact latent representations of channel matrices. However, most existing solutions remain highly task-specific and configuration-dependent~\cite{9796036},~\cite{9134938}.  The representations learned by these models are tightly coupled to the particular antenna array dimensions and user configurations encountered during training. As a result, a model trained for one deployment scenario often exhibits substantial performance degradation when applied to another.

\gls{cfm} provides an alternative by learning transferable channel representations from large-scale \gls{csi} datasets. Rather than optimizing for a single network configuration, \gls{cfm} aims to capture representations that generalize across different deployment scenarios and radio configurations. This transferability is valuable for \gls{fdd} precoding and beamforming, where operators must support multiple configurations with a single model.

As an example, WiFo-CF~\cite{11479626} is a \gls{cfm} designed for \gls{csi} feedback. It uses heterogeneous configurations—varying channel dimensions, feedback rates, and data distributions, and employs a multi-user, multi-rate self-supervised pretraining strategy and a Mixture of Shared and Routed Expert (S-R MoE) architecture. Evaluations demonstrate that WiFo-CF achieves superior performance on both in-distribution and out-of-distribution data across simulated and real-world scenarios, highlighting the potential of \gls{cfm} to provide scalable \gls{csi} feedback solutions across diverse network configurations.

WiFo-CF handles heterogeneous \gls{csi} dimensions and feedback settings, but real deployments also vary site-to-site across propagation environments. To address this challenge, SiFo~\cite{zhao2026sifowirelessfoundationmodel} proposes a \gls{cfm} framework for low-overhead site-specific \gls{csi} feedback. SiFo pretrains a \gls{csi} feedback model across multiple source sites and adapts it to a target site through lightweight calibration using a small set of reference users. Specifically, the target site provides low-dimensional reference signal received power (RSRP) fingerprints from a small number of users, which are used as site-specific guidance to calibrate the pretrained model without requiring full-scale retraining on target-site \gls{csi} data. By combining transferable representations learned during pretraining with this low-dimensional site-specific information, SiFo reduces the data overhead required for adaptation while maintaining accurate \gls{csi} representation. Evaluations across ten city-scale scenarios show that SiFo achieves higher \gls{csi}-capture efficiency than site-specific learning approaches and approaches the performance of the high-overhead 3GPP NR Type-II feedback framework while relying only on lightweight online measurements.

Complementing these \gls{csi} feedback frameworks, WiFo-E~\cite{wen2026wifoescalablewirelessfoundation} extends the foundation-model paradigm to end-to-end \gls{fdd} precoding. WiFo-E learns compact and transferable \gls{csi} representations that generalize across varying numbers of antennas and users. By jointly addressing \gls{csi} compression, \gls{csi} reconstruction, and precoder design within a unified framework, WiFo-E demonstrates superior scalability compared with conventional task-specific models.

These results suggest that \glspl{cfm} can reduce the need to retrain and redeploy models whenever network configurations change. This makes managing the \gls{ai} lifecycle easier in future wireless systems. The results also show that \gls{cfm} can overcome the scalability and generalization problems of conventional \gls{csi} feedback methods, which makes them well suited for diverse 5G-Advanced and future 6G deployments.

Although recent \gls{cfm}-based \gls{csi} feedback frameworks have shown promising results, several challenges remain:
\begin{enumerate}
    \item \textbf{Limited multi-user transmission scenarios.} Existing \gls{cfm}-based methods mainly focus on single-user settings, while future networks will involve more complex scenarios such as Multi-User-\gls{mimo}, Cell-Free Massive \gls{mimo}, and Multi-Transmission Reception Point communications. Extending \glspl{cfm} to efficiently represent and compress jointly structured multi-user \gls{csi} while preserving the information required for joint precoding remains an open challenge.

    \item \textbf{Robustness in dynamic environments.} Real-world deployments must handle quickly changing channel conditions caused by user mobility, signal blockage, and other environmental changes. This calls for future research on more robust and adaptive \gls{csi} feedback and precoding frameworks.

    \item \textbf{Limited availability of real-world training data.} The scarcity of large-scale and diverse real-world \gls{csi} datasets limits \gls{cfm} training and evaluation. Collecting such datasets at sufficient scale remains challenging due to the cost and complexity of real-world measurements.
\end{enumerate}
 
\subsection{End-to-End Communication Systems}
\label{subsec:e2e}
 
Although recent \glspl{wpfm} have demonstrated the ability to support multiple downstream tasks, most existing models remain focused on a collection of related functions, such as channel estimation, channel prediction, and \gls{csi} feedback. This raises a broader question: can the \gls{wpfm} paradigm be extended beyond individual communication blocks to learn representations of the communication system as a whole?

A practical communication system consists of multiple interconnected blocks, including channel estimation, signal detection, \gls{csi} feedback, precoding, beam management, modulation and coding, and resource allocation. The objective of an end-to-end communication system \gls{wpfm} is to learn representations that capture the interactions among these blocks and the underlying communication process. Such representations could then be adapted to perform the functionality of any block when required, reducing the need to design and train dedicated models for each communication task. This capability is important in dynamic wireless environments, where the performance of individual blocks may vary across communication conditions.

An early step toward this vision is WiFo-E~\cite{wen2026wifoescalablewirelessfoundation}, a scalable \gls{wpfm} for end-to-end \gls{fdd} precoding. Rather than optimizing individual stages of the channel estimation, \gls{csi} feedback, and precoding separately, WiFo-E jointly addresses all of them through a shared representation. By spanning multiple stages of the communication process, WiFo-E demonstrates how \gls{wpfm} can learn transferable representations that support separate communication blocks.

A complementary step toward end-to-end communication-system \gls{wpfm} is \cite{buffelli2025towards}. The authors propose a framework that jointly models the parameters of multiple communication blocks, including transmission rank, precoder selection, Doppler spread, and delay-profile characteristics. This enables the model to capture dependencies among different communication-system blocks, demonstrating that \glspl{wpfm} can learn representations extending beyond tasks of a single communication block.

Despite these results, end-to-end \glspl{wpfm} are still in their early stages. Two main directions stand out for future work:

\begin{enumerate}
    \item \textbf{Toward a unified framework.} Current models support only a limited set of communication blocks. A key goal is to extend them into a single, unified framework that can handle a much wider range of functionalities.

    \item \textbf{Balancing generality and specialization.} As more communication tasks are added to a shared model, making its representations more general may reduce performance on individual tasks. Understanding this trade-off and designing effective ways to balance shared knowledge against task specialization, will be essential for future end-to-end \glspl{wpfm}.

    \item \textbf{Continual and online optimization.} Communication conditions and network requirements evolve over time, requiring end-to-end \glspl{wpfm} to continuously adapt their representations and decisions. Efficient online adaptation with low computational and signaling costs remains an open challenge.
\end{enumerate}

\subsection{Large-Scale AI Integration and Distributed Deployment}
\label{subsec:6gfederated}

The increasing scale and capabilities of \glspl{wpfm} are reshaping the relationship between \gls{ai} and telecommunication networks. Traditionally, wireless networks have served primarily as communication infrastructures that transport data generated by \gls{ai} applications. In contrast, emerging 6G visions increasingly consider \glspl{wpfm} as native network entities that participate directly in communication, sensing, and management processes. This transition introduces a new research challenge: enabling the seamless integration of large-scale \gls{ai} systems within telecommunication infrastructures.

Several recent studies have explored this integration. Shahid et al.~\cite{shahid2025large} provide a roadmap for integrating large-scale AI into telecommunication infrastructures, covering model governance, scalability, and regulatory considerations. Similarly, Maatouk et al.~\cite{maatouk2024large} examine the use cases of \glspl{llm} that can be readily 
implemented in the telecom industry, streamlining tasks, such as anomaly resolution and technical 
specification comprehension. Extending this vision toward future wireless systems, Chen et al.~\cite{10579546} provide a comprehensive discussion as well as some in-depth prospects on the demand, design, and deployment aspects of \glspl{wpfm}. They opine that \glspl{wpfm} will be a recipe for the 6G wireless networks to build high-efficient, sustainable, versatile, and extensible wireless intelligence for numerous promising visions. Likewise, Shoaib et al.~\cite{10683090} highlight the strong alignment between \gls{wpfm} capabilities and the intelligence requirements of future 6G architectures. The authors discuss how \glspl{wpfm} can enhance 6G communications and vice versa, outlining applications such as the Internet of Vehicles and the Metaverse. These studies indicate that future telecommunication networks are expected to evolve to take advantage of large-scale \gls{ai} models as integral components of network operation. This shift has the potential to enable more autonomous, adaptive, and context-aware wireless systems capable of exploiting knowledge across multiple network functions. 

However, realizing this vision introduces several significant practical challenges. First, \glspl{wpfm} often contain millions of parameters, resulting in substantial computational and memory requirements during both training and inference. As the scale of these models continues to increase, the computational resources required to support them may exceed the capabilities of individual devices.

Furthermore, future \glspl{wpfm} are increasingly evolving toward multi-modal architectures that jointly exploit different types of data. These modalities are often generated and stored across multiple distributed sources, such as \glspl{ue}, \glspl{bs}, sensors, cameras, and UAVs. Consequently, training and operating multi-modal \glspl{wpfm} may require the collection of large volumes of data distributed throughout the network. Such a process introduces substantial communication overhead and consumes wireless resources. As the volume of data and the number of participating devices continue to increase, data transmission may become a major bottleneck for large-scale \gls{ai} deployment in wireless networks.

In addition to communication overhead, centralized processing raises important privacy and security concerns. Many of the data sources involved in future \gls{ai}-native wireless systems may contain sensitive user information, location data, mobility patterns, or network operational information. Transferring such data to centralized servers for model training or inference may conflict with privacy requirements.

Beyond communication and privacy considerations, centralized deployment may also negatively affect the timeliness of \gls{ai}-assisted wireless decisions. Many wireless functions, including channel adaptation, beam management, and resource allocation, operate under strict latency constraints and rely on rapidly changing network information. When raw data must first be transmitted to centralized computing resources for inference and the resulting decisions subsequently returned back to users or network entities, the associated end-to-end delay may cause the inferred information to become outdated before it can be utilized. This issue becomes particularly critical in highly dynamic wireless environments where channel conditions, user locations, and network states evolve rapidly. Consequently, the effectiveness of \gls{ai}-assisted wireless systems depends not only on model accuracy but also on the ability to perform learning and inference with sufficiently low latency and up-to-date information.

To address these challenges, recent research has increasingly focused on distributed deployment strategies for foundation models. Among these approaches, federated learning has emerged as a promising framework for training and adapting \glspl{wpfm} without requiring centralized access to raw data. Chen et al.~\cite{10558823} explore the extent to which \glspl{wpfm} are suitable for federated learning over wireless networks; the authors also discuss multiple new paradigms for realizing future intelligent networks that integrate \glspl{wpfm} and federated learning. Their work further highlights the potential of federated learning to support model adaptation.

Beyond distributed training, edge deployment enables \gls{wpfm} inference and adaptation to be performed closer to end users and data sources. To this end, Wu et al.~\cite{10558820} propose a device-edge fine-tuning framework for providing efficient and privacy-preserving fine-tuning services at the 6G network edge. By maintaining task-specific adapters on user devices, the proposed framework reduces data exposure while preserving personalization capabilities and adaptation to local environments.

For highly resource-constrained devices, model complexity itself becomes a major deployment bottleneck. Addressing this challenge, Cheraghinia et al.~\cite{cheraghinia2025lightweightfoundationmodelwireless} present a lightweight \gls{wpfm} based on simple \gls{mlp} encoders; the model achieves near-large-model performance with a fraction of the parameters. The proposed framework demonstrates the feasibility of deploying \glspl{wpfm} on low-power edge hardware with limited computational and memory resources.

In addition to model compression, collaborative edge-computing architectures have also been explored to support large-scale foundation-model deployment. Zeng et al.~\cite{10.1109/MWC.004.2300479} propose collaborative edge training, a training mechanism that orchestrates a group of trusted edge devices 
as a resource pool for expedited, sustainable big \glspl{ai} model training at the edge. By distributing workloads across edge infrastructures, the framework reduces reliance on centralized computing, thereby alleviating backhaul communication bottlenecks, lowering datacenter energy consumption, and improving privacy by minimizing the transmission of raw user data to remote processing facilities.

Together, these approaches provide a practical pathway for operating \glspl{wpfm} in \gls{ai}-native 6G networks while satisfying the strict requirements of wireless systems in terms of latency, privacy, scalability, and resource efficiency.

\subsection{Computational Aspects of Telecommunications Tasks}
\label{subsec:compute_telecom}

Deploying foundation-model-based inference within a real-time telecommunications pipeline fundamentally differs from running a model in a cloud data center. Every task in the radio stack faces strict latency deadlines dictated by communication standards, and the processing hardware typically consists of purpose-built silicon with tightly bounded memory resources. 

\subsubsection{Implications for Model Size}

Computational constraints vary significantly across tasks and within the UE--gNB--edge--cloud hierarchy:

\paragraph{Channel estimation and equalization}
These tasks reside in the innermost loop of the baseband processing pipeline and must be completed within a fraction of an OFDM symbol duration (e.g., $\sim\!70\,\mu$s at 15\,kHz subcarrier spacing). On a UE baseband ASIC with 1--4\,MB of SRAM, only models with fewer than approximately \textbf{200\,K--500\,K parameters} (quantized to INT8 or below) are feasible. At the gNB, the allowable budget is somewhat larger---up to \textbf{1--5\,M parameters}---due to increased SRAM and pipelined operation across antenna ports. This precludes direct deployment of large foundation models; task-specific heads distilled from a pretrained backbone are therefore more practical.

\paragraph{FDD precoding and CSI feedback}
CSI computation at the UE, which consists of encoding channel estimates into a compact codeword, is restricted by tight latency and memory constraints, typically a few milliseconds and 2--4\,MB of memory. This hardware asymmetry necessitates a lightweight encoder at the UE (generally $<$1\,M parameters), while the decoder at the gNB can accommodate greater complexity (often 5--10\,M parameters). For example, Cheraghinia et al.~\cite{cheraghinia2025lightweightfoundationmodelwireless} proposed a lightweight real-time model with approximately 21\,K parameters for UE-side technology recognition and LoS detection; foundation models such as WiFo-E~\cite{wen2026wifoescalablewirelessfoundation} have implemented a precoding engine at the gNB with approximately 11\,M parameters.

\paragraph{End-to-end communication}
A unified neural Tx/Rx chain must process each transport block within the slot-level latency budget. For URLLC (1\,ms end-to-end), neural processing can occupy only a fraction of this window---typically $200$--$500\,\mu$s. This constrains the combined Tx and Rx model to roughly \textbf{0.5--2\,M parameters} on current baseband hardware, assuming INT8 inference at $\sim\!1$\,TOPS throughput typical of advanced DSPs.

\paragraph{Semantic and goal-oriented communications}
With latencies of 10--100\,ms and access to application-processor NPUs capable of 10--30\,TOPS (e.g., Qualcomm Hexagon, Apple Neural Engine), model sizes of \textbf{10--100\,M parameters} are feasible on the UE side. Edge servers equipped with data-center GPUs (NVIDIA A100/H100) can execute \textbf{hundreds of millions to billions} of parameters within the same latency constraints.

\paragraph{Federated and edge fine-tuning}
Fine-tuning tasks are latency-tolerant (seconds to minutes), thus the primary constraint is memory. Full fine-tuning of a 100,M-parameter model requires $\sim\!1.2$\,GB of optimizer states in FP16 mixed-precision training. Parameter-efficient methods (LoRA, adapters) can reduce trainable parameters by $10$--$100\times$, enabling practical fine-tuning of \textbf{100, M+ parameter} backbones on edge GPUs with 8--16\,GB memory and even on UE NPUs for adapter-only updates.

\paragraph{6G large-scale AI integration}
Network-level optimization (e.g., RAN slicing, traffic steering) operates on centralized or near-real-time RAN Intelligent Controllers (RICs) equipped with server-grade GPUs. Here, models with \textbf{hundreds of millions to billions of parameters} are deployable, and the challenge shifts from raw compute to orchestration, versioning, and model coordination.


\section{Localization Applications}
\label{sec:localization}

\begin{figure*}
    \centering
    \includegraphics[width=\linewidth]{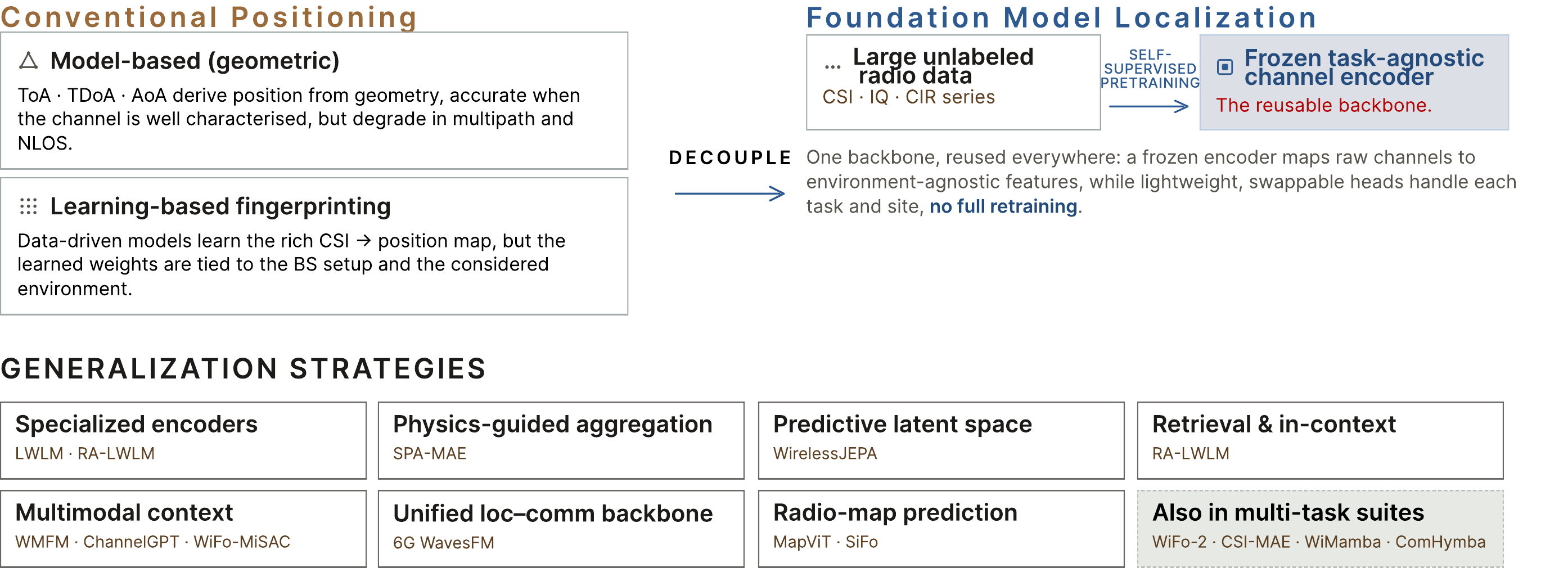}
        \caption{Transition from conventional geometric and fingerprint-based positioning to foundation-model localization.}

    \label{fig:localization_app}
\end{figure*}

Precise positioning is an important service in 6G for autonomous systems, \ac{xr}, the low-altitude economy, and location-aware network optimization~\cite{11315904}. There are two main paradigms. \textbf{Model-based} methods exploit geometric relationships, such as time-of-arrival, time-difference-of-arrival, and angle-of-arrival, to calculate the \gls{ue} position; they achieve high accuracy when the propagation environment is well characterized but degrade severely in multipath-rich and \gls{nlos} conditions. \textbf{Learning-based} fingerprinting instead assigns the channel-to-position mapping to data-driven models that learn the rich features embedded in \gls{csi}. The defining limitation of learning-based localization is that the learned parameters are tightly coupled to the \gls{bs} configuration (antenna count, bandwidth, orientation) and the propagation environment of the training environment. Consequently, every new deployment requires recollecting fingerprints and retraining the model, which introduces \textbf{distribution shift across sites, configurations, and environments}, known as the central challenge of the field. The growing importance of accurate positioning is also reflected in the 3GPP Release-18 and Release-19 study items on AI/ML-assisted positioning~\cite{lin2023overview3gppstudyartificial}.

\glspl{wpfm} address this challenge by learning transferable radio representations from large unlabeled data and then decoupling environment-invariant representation learning from environment-specific adaptation, as illustrated in Fig.~\ref{fig:localization_app}. This section is therefore organized around the \textbf{generalization strategy} that each model adopts to avoid retraining.

\subsection{From Fingerprinting to Foundation Models}
\label{subsec:rf_positioning}
 
Fingerprinting requires large environment surveys to build labeled databases, and the resulting models do not transfer across environments. Foundation-model approaches instead learn generalized radio-environment representations that can be reused with minimal environment-specific effort. \cite{11264506} develops a unified \gls{wpfm} that jointly handles wireless technology recognition and localization directly from raw \gls{iq} and \gls{cir} time series, demonstrating that a single self-supervised backbone can serve both tasks while reducing ranging \gls{mae} by up to 50\% and detecting \gls{los}/\gls{nlos} conditions with high accuracy. More generally, the large wireless model (LWM)~\cite{alikhani2024largewirelessmodellwm} learns cross-scenario channel representations that can be used as a feature extractor for positioning. These works establish the key basis of the section that a frozen, task-agnostic channel encoder provides a reusable backbone on top of which lightweight, task-specific localization can be built.

\subsection{Specialized Localization Foundation Models}
\label{subsec:specialized_loc}

Beyond reusing communication-centric backbones, a line of work designs \glspl{wpfm} specifically for positioning. The Large Wireless Localization Model (LWLM)~\cite{pan2025largewirelesslocalizationmodel} introduces a channel encoder tailored to localization that surpasses both model-based and supervised baselines. Building directly on LWLM, RA-LWLM~\cite{pan2026ralwlmretrievalaugmentedincontextlocalization} adapts the retrieval-augmented generation and in-context learning paradigms from \ac{nlp} to wireless positioning. A frozen \gls{wpfm} encoder maps \gls{csi} to environment-agnostic representations; a retrieval module finds nearest-neighbor references in a per-environment fingerprint database; and a \gls{moe}-based in-context localization module produces the final position as a weighted mixture. The main idea is to keep environment-specific information in a separate, replaceable database rather than adding it into the model's weights. To adapt to a new environment, you just update the database without needing to retrain the model. In experiments, RA-LWLM achieves a median localization error of 0.53\,m on unseen environments, only 8.2\% degradation from seen environments, with zero retraining and nearly matching per-environment retrained baselines.

\subsection{Generalization Strategies}
\label{subsec:cross_env_loc}

The surveyed models pursue distribution-shift robustness through several complementary mechanisms.

\textbf{Physics-guided global aggregation:} SPA-MAE~\cite{chen2026spamaephysicsguidedcsifoundation} augments masked reconstruction with structure-aware and parameter-aware auxiliary objectives and aggregates propagation-aware information. This yields strong cross-city positioning, achieving 2--3$\times$ lower mean distance error than LWM across six unseen cities (e.g., 12.37\,m versus 28.05\,m in Denver) with only 0.33\,M parameters.

\textbf{Predictive and latent representations:} WirelessJEPA~\cite{chu2026wirelessjepamultiantennafoundationmodel} performs spatio-temporal latent prediction over multi-antenna arrays, learning representations that are robust to environmental change and directly applicable to fingerprinting and angle-of-arrival estimation. Operating in latent rather than signal space makes the learned features less sensitive to environment-specific low-level details.

\textbf{Retrieval and in-context adaptation:} RA-LWLM (Section~\ref{subsec:specialized_loc}) exemplifies training-free cross-environment adaptation by supplying environment-specific knowledge externally at inference time.

\textbf{Multimodal context:} Several models reduce scenario sensitivity by fusing \gls{csi} with environmental side information. \cite{10971878} combines \gls{csi} with geographic maps and \gls{bs}/\gls{ue} positions to mitigate scenario-generalization issues, while ChannelGPT~\cite{11180834} injects position and environmental context into channel modeling. WiFo-MiSAC~\cite{liu2026wifomisacwirelessfoundationmodel} pushes this to zero-shot multimodal positioning across unseen links, and the Wireless Multimodal Foundation Model (WMFM)~\cite{farzanullah2025wirelessmultimodalfoundationmodel} jointly learns vision and communication representations for \gls{isac} tasks including localization.

\textbf{Unified-backbone evidence:} A number of multi-task \glspl{wpfm} report positioning as a downstream task and confirm that a shared backbone generalizes across environments. Notably, 6G~WavesFM~\cite{11131142} attains roughly half the positioning error of a supervised baseline while sharing 80\% of its parameters across four tasks, and WiFo-2~\cite{liu2025foundationmodelintelligentwireless}, CSI-MAE~\cite{jiang2026csimaemaskedautoencoderbasedchannel}, WiMamba~\cite{raviv2026wimambalinearscalewirelessfoundation}, and ComHymba~\cite{yang2026comhymbalowcomplexitydomaininformedfoundation} all include localization within their multi-task suites.

\subsection{Radio-Map and Propagation-Aware Prediction}
\label{subsec:radiomaps}

Distinct from \gls{ue} coordinate estimation, a related task predicts the spatial distribution of radio quality over an environment. Radio quality maps encode signal strength, interference, and coverage, and serve network planning, beam management, and localization aiding. MapViT~\cite{hsu2026mapvittwostagevitbasedframework} proposes a two-stage \gls{vit} framework for real-time radio-quality-map prediction in dynamic environments, using depth maps as input and exploiting self-attention to model long-range reflections that \glspl{cnn} with local receptive fields systematically miss. We emphasize that this is propagation/coverage mapping rather than UE positioning; it nonetheless complements localization by characterizing the environment in which positioning operates. The same fingerprint-to-environment principle appears in SiFo~\cite{zhao2026sifowirelessfoundationmodel}, which exploits low-dimensional \gls{rsrp} fingerprints for site-specific adaptation, illustrating the tight link between positioning fingerprints and channel representation.

\subsection{Towards a Unified Localization and Communication Backbone}
\label{subsec:unified_loc}

The convergence of communication and localization functions in 6G, such as position-aided beamforming and the duality between channel representation and positioning, motivates unified foundation models that serve both goals from a single backbone. The unified model \cite{11264506} and the multi-task backbones represent early steps toward this vision, training on combined channel-position data and exposing localization as one of several downstream heads. A practical consequence, analyzed in Section~\ref{subsec:compute_localization}, is that a dual-purpose backbone must respect the stricter communication latency budget while a lower-rate localization head operates alongside it. This unification ultimately points toward \gls{isac}, where positioning becomes one instance of the broader environmental-perception capability examined in Section~\ref{sec:sensing}.

\subsection{Computational Aspects of Localization Tasks}
\label{subsec:compute_localization}

Localization tasks are generally less latency-critical than inner-loop baseband processing, but they still impose meaningful real-time constraints, especially in safety-critical applications such as autonomous driving and industrial robotics. 

\subsubsection{Implications for Model Size}

\paragraph{RF-based positioning for pedestrian navigation}
At 1--10\,Hz with a latency budget of 50--100\,ms, the inference requirements are modest. Modern smartphone NPUs (10--30\,TOPS) can execute models with \textbf{10--50\,M parameters} (INT8) well within this window, leaving room for concurrent sensor fusion with IMU and GNSS data. Network-side positioning, performed on the Location Management Function (LMF) server in 5G, has essentially no model-size constraint beyond server memory (typically 16--64\,GB). Models with \textbf{50--200\,M parameters} are readily deployable.

\paragraph{High-rate vehicular and industrial tracking}
At 10--100\,Hz, the per-inference budget drops to 10--50\,ms. Edge inference platforms such as NVIDIA Orin (275\,TOPS INT8, 32\,GB unified memory) can handle models of \textbf{20--80\,M parameters} at this rate. The critical consideration is not a single inference but the sustained throughput when tracking multiple targets concurrently; for $N$ targets, the effective per-target budget is divided by $N$ unless batched inference is used.

\paragraph{Cross-environment adaptation}
Adapting a pretrained localization backbone to a new site is an offline or infrequent process, so model-size constraints are relaxed. This is where full-size foundation models (\textbf{100\,M--1\,B parameters}) can be employed directly. Parameter-efficient fine-tuning methods (LoRA, adapters) allow the adaptation to complete within minutes on a single GPU using a few hundred labeled calibration samples.

\paragraph{Real-Time Beam Prediction}
Geo-aware beam prediction is a latency-critical task run on gNB-connected edge servers to accelerate beam alignment. Architectures like the ViT-based Map-ViT~\cite{hsu2026mapvittwostagevitbasedframework} are engineered for this real-time environment, achieving inference latencies as low as 0.17\,ms. This approach offloads all computational complexity from the UE to the network edge from 5M-parameter model to 86M parameters.

\paragraph{Unified localization--communication backbone}
The dual-purpose backbone must respect the stricter of the two latency constraints (sub-ms for communication). In practice, this motivates a shared encoder of \textbf{1--5\,M parameters} deployed on the gNB BBU with separate lightweight heads: one for real-time channel tasks and one for localization operating at a lower rate. The localization head may additionally offload to a co-located accelerator when higher accuracy is needed.

\section{Sensing Applications}
\label{sec:sensing}

 \begin{figure*}
     \centering
     \includegraphics[width=\linewidth]{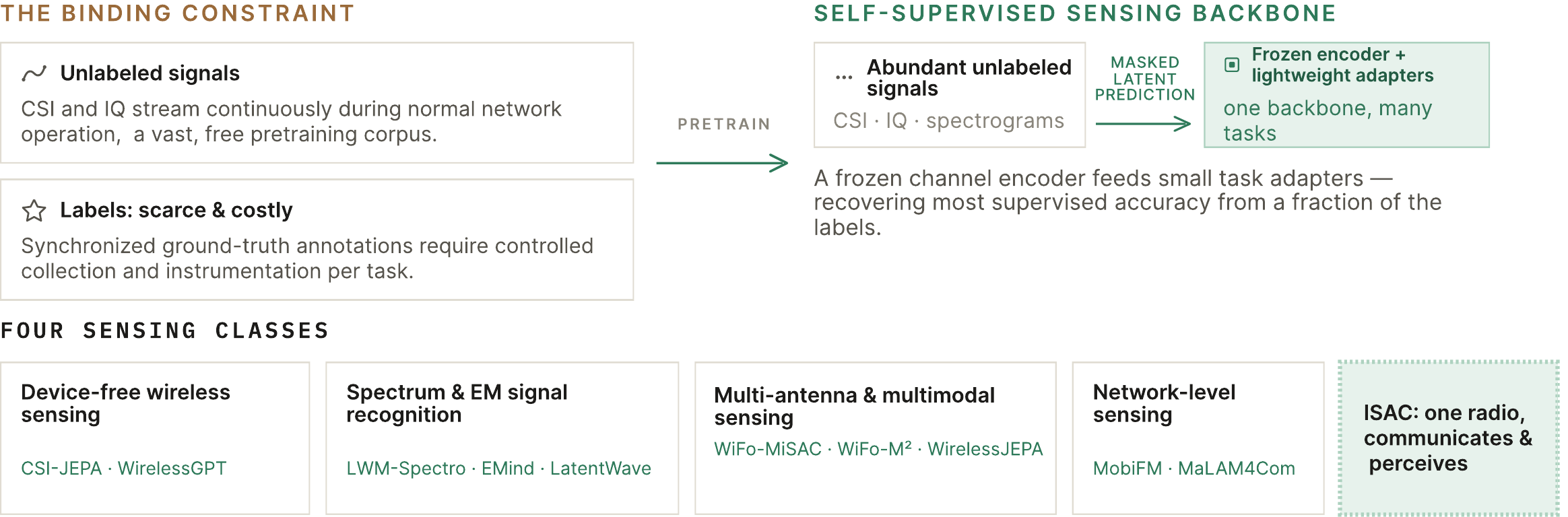}
    \caption{Foundation-model approach to wireless sensing.}
     \label{fig:sensing_app}
 \end{figure*}
 
Sensing is the perception counterpart to communication in 6G, the same radio infrastructure that transports data can also infer the state of the physical environment. \gls{isac} is the unifying paradigm for this convergence, in which a single hardware platform, spectrum allocation, and waveform jointly serve communication and sensing~\cite{kaushik2023integratedsensingcommunications6g, 11404239}. Within this vision, the intelligent fusion of communication and multimodal sensing has been formalized as the Synesthesia of Machines~\cite{10330577}. Sensing tasks, however, span several genuinely different problem classes that differ in what is being sensed and which physical signal is used. We organize this section accordingly into four classes, device-free \gls{csi}-based sensing, spectrum and electromagnetic signal recognition, multi-antenna and multimodal environmental sensing, and network-level sensing, and then synthesize how they converge under \gls{isac}, as summarized in Fig.~\ref{fig:sensing_app}. A recurring theme distinct from communication is that the binding constraint for sensing is \emph{label scarcity}: unlabeled signals are abundant during normal operation, whereas synchronized ground-truth annotations are expensive to collect, which makes self-supervised pretraining especially valuable.

\subsection{Device-Free Wireless Sensing}
\label{subsec:devicefree}

Device-free sensing analyzes the perturbations that human motion, breathing, presence, and environmental change induce on wireless channels, enabling perception without requiring targets to carry any device. \gls{csi} from commodity Wi-Fi is the dominant modality, increasingly standardized through IEEE~802.11bf. CSI-JEPA~\cite{luo2026csijepafoundationrepresentationsubiquitous} is a dedicated foundation model for this setting. Wi-Fi \gls{csi} phase measurements are often corrupted by hardware impairments such as carrier frequency offset, sampling frequency offset, packet detection delay, and synchronization errors; therefore, CSI-JEPA uses the amplitude component as its input. It tokenizes the \gls{csi} amplitude matrix into two-dimensional patches spanning multiple time samples and subcarriers. These patches are treated as tokens, and the model learns reusable representations by predicting the latent features of masked patches from the surrounding context. Its channel variation-aware masking samples predictive targets from regions with stronger temporal and subcarrier-domain dynamics, focusing the pretext task on sensing-informative channel variations. Across seven real-world tasks (fall detection, human activity recognition, breathing detection, room-level localization, motion-source recognition, proximity recognition, and user identification), a frozen encoder with only 0.067\,M trainable adapter parameters achieves up to 98\% labeled-data savings over supervised baselines. A central finding is that \gls{jepa}-style latent prediction outperforms \gls{mae-model}-style raw reconstruction for sensing, because reconstruction over-weights low-level amplitude detail that is not the most discriminative factor for downstream tasks. WirelessGPT~\cite{11278185} similarly supports human activity recognition among its downstream tasks, confirming that a self-supervised channel backbone transfers to device-free sensing under limited labels.

\subsection{Spectrum Sensing and Electromagnetic Signal Recognition}
\label{subsec:spectrum}

Automatic modulation classification, signal detection, technology recognition, and emitter identification are canonical spectrum-sensing tasks, and foundation models pretrained on diverse \gls{rf} environments naturally capture inter-technology spectral patterns. LWM-Spectro~\cite{kim2026lwmspectrofoundationmodelwireless} operates on baseband signal spectrograms, applying masked and contrastive Transformer pretraining with protocol-specialized \gls{moe} experts to achieve cross-standard signal recognition across WiFi, LTE, and 5G. EMind~\cite{luo2025emindfoundationmodelmultitask} broadens this to electromagnetic signal understanding over raw \gls{iq}, handling signal classification, modulation recognition, and emitter identification within a single backbone, and introducing multi-signal packing to cope with the large variation in signal length across heterogeneous sources. The unified \gls{wpfm} of Cheraghinia et al.~\cite{11264506} couples technology recognition with localization (Section~\ref{subsec:rf_positioning}), and LatentWave~\cite{mohamed2026latentwavejepapretrainingwireless} adds \gls{rf} classification to its \gls{jepa} downstream suite. Beyond classification, foundation models pretrained on normal spectral distributions can flag deviations indicative of interference or abnormal emissions, extending self-supervised representations from recognition to proactive anomaly detection and spectrum management. Self-supervised pretraining on large signal databases also captures subtle hardware-induced transmitter imperfections, such as I/Q imbalance, phase noise, and power-amplifier nonlinearities, enabling \gls{rf} fingerprinting for device identification and authentication.

\subsection{Multi-Antenna and Multimodal Environmental Sensing}
\label{subsec:env_sensing}

The richest sensing class uses multi-antenna \gls{csi}, often fused with radar and visual modalities, to perceive the physical environment, and it is here that sensing and communication merge most tightly. WiFo-MiSAC~\cite{liu2026wifomisacwirelessfoundationmodel} is a task-agnostic foundation model for multimodal sensing-communication integration that tokenizes \gls{csi}, \gls{fmcw} radar (as range-angle and range-velocity maps), and \gls{bev} maps into a unified space and employs a shared-specific disentangled \gls{moe} to decouple modality-shared from modality-specific representations, avoiding cross-modal interference. It demonstrates robust few-shot adaptation, seamless integration of new modalities, and graceful degradation under missing modalities (a 1.33\,dB drop versus 5.11\,dB for task-specific models). WiFo-M$^2$~\cite{zhang2026wifom2plugandplaymultimodalsensing} adopts a plug-and-play approach in which LiDAR and image sensing modules empower communication through a shared backbone, using a temporal sequence-to-sequence model to extrapolate channel-aligned sensing features and fill the gaps between low-rate sensor frames. WirelessGPT~\cite{11278185} extends environmental perception to environment reconstruction and object tracking, and WirelessJEPA~\cite{chu2026wirelessjepamultiantennafoundationmodel} learns compact spatio-temporal representations from massive-\gls{mimo} arrays for angle-of-arrival estimation, modulation classification, and \gls{rf} fingerprinting. The multi-task backbone 6G~WavesFM~\cite{11131142} likewise treats human-activity sensing and \gls{rf} classification as heads on a shared \gls{vit}. The vision-and-communication WMFM~\cite{farzanullah2025wirelessmultimodalfoundationmodel} reinforces the same direction for \gls{isac}. As already noted for multimodal tokenization in Section~\ref{subsec:tax_modality}, the principal open problem in this class is the same temporal misalignment between fast-varying \gls{rf} channels and slower visual sensors, at the sensing-fusion level,  cross-modal alignment mechanisms are required to fuse asynchronous multi-rate observations, for example through the sequence-to-sequence feature extrapolation of WiFo-M$^2$.

\subsection{Network-Level Sensing}
\label{subsec:traffic}

Distinct from physical-layer perception, network-level sensing predicts traffic loads, mobility patterns, and resource demands for proactive network management. We include it here for completeness while noting that it operates on aggregated telemetry rather than radio signals. MobiFM~\cite{11298194} is a foundation model for mobile data forecasting that generalizes across cities and time horizons, addressing the heterogeneity inherent in large-scale network telemetry. Collaborative settings extend this to multiple cooperating agents: MaLAM4Com~\cite{11303197} proposes a multi-agent cooperative large-AI-model framework in which knowledge is transferred across distributed nodes via federated distillation and low-dimensional embeddings, enabling collaborative channel estimation and beam prediction without centralizing raw data. Although MaLAM4Com is primarily a distributed-deployment and cooperative-communication framework (discussed alongside the large-scale integration approaches of Section~\ref{subsec:6gfederated}), its architecture also provides building blocks for autonomous, collaborative spectrum-management systems in which distributed agents coordinate to monitor and protect shared spectral resources in real time.

\subsection{ISAC as the Convergence Point}
\label{subsec:isac}

The four classes above converge under \gls{isac}: a single radio platform that simultaneously communicates and perceives. Their unification is enabled by a deeper structural fact, namely that \gls{csi} and \gls{iq} are simultaneously the input for communication, the fingerprint for localization, and the perturbation signal for sensing. This is why the unified backbones, such as WirelessGPT, WiFo-MiSAC, and 6G~WavesFM, can serve sensing, localization, and communication from one pretrained model, and why localization can be viewed as a special case of sensing in which the estimated environmental variable is the user position. Environment-aware sensing also feeds back into communication: the channel foundation models of Section~\ref{subsec:channel} that internalize the causal link between scene geometry and signal propagation, together with \gls{eic}~\cite{ZHANG2026186} and wireless digital twins~\cite{10148936, 10742564}, depend on exactly the multimodal environmental perception developed here. The shared open challenges, asynchronous multi-rate fusion and the absence of a discrete sensing vocabulary, are common to both this section and the multimodal channel modeling of Section~\ref{sec:telecom}, underscoring that \gls{isac} is best treated as a convergence of the two rather than as a separate silo.

\subsection{Computational Aspects of Sensing Tasks}
\label{subsec:compute_sensing}

Sensing tasks span a broad spectrum of computational profiles, ranging from latency-critical radar-communication processing in ISAC to latency-tolerant network traffic forecasting. 

\subsubsection{Implications for Model Size}

\paragraph{Integrated sensing and communication}
ISAC processing must synchronize with the slot-level cadence of the communication system. For a pulse repetition interval (PRI) of 0.5--1\,ms, the sensing inference share is at most a few hundred microseconds. On a gNB-class FPGA, this limits neural sensing models to approximately \textbf{0.5--2\,M parameters} at INT8 precision.

\paragraph{Spectrum sensing and signal recognition}
In 5G New Radio and future 6G networks, sensing and signal-recognition functions are often constrained by the underlying radio numerology and measurement periodicity, resulting in latency budgets ranging from sub-millisecond to tens of milliseconds. Cognitive radio standards (e.g., IEEE~802.22) similarly mandate sensing decisions within a few milliseconds. At invocation intervals between 1--100\,ms, an FPGA-based sensor with 1--16\,MB of on-chip memory accommodates models of \textbf{100\,K--1\,M parameters}. Edge-based deployments can relax this to \textbf{5--50\,M parameters}, enabling richer classifiers such as LWM-Spectro~\cite{kim2026lwmspectrofoundationmodelwireless} with 9.2\,M parameters. Monitoring-grade applications with 100\,ms latency budgets allow for substantially larger models.

\paragraph{Multi-antenna spatio-temporal sensing}
Processing the full spatio-temporal data cube from a massive MIMO array (64--256 antennas) is computationally intensive even for traditional algorithms. Neural models like WirelessJEPA~\cite{chu2026wirelessjepamultiantennafoundationmodel} must be optimized for efficient antenna-domain processing. At a 1--10\,ms cadence on a gNB BBU, models of \textbf{2--10\,M parameters} are practical on FPGA/ASIC, while GPU offloading extends feasibility to \textbf{20--80\,M parameters}. Factorized architectures that separately address spatial (antenna) and temporal (slot) domains can significantly reduce effective model size without sacrificing capacity.

\paragraph{Mobile data and traffic forecasting}
With invocation every 5-15 minutes and access to server-class hardware, traffic forecasting is the least constrained sensing task. For example, MobiFM-class models~\cite{11298194} are deployed at 13\,M--214\,M parameters. The primary constraint is maintaining and updating such models across hundreds of cells in large-scale networks.

\paragraph{Multi-agent cooperative sensing}
In collaborative sensing, each agent (base station, UAV, vehicle) operates its own local model and shares compressed embeddings with peers. Per-agent models must fit the device compute envelope (\textbf{1--20\,M parameters} on edge FPGA/GPU), while aggregation models at the coordinating node can be somewhat larger (\textbf{10--50\,M parameters}). Communication overhead for exchanging embeddings adds to total latency and should be jointly designed with model architecture, as highlighted by~\cite{11303197}.

\section{Bridging the Gap: Model Compression and Efficient Deployment}
\label{sec:compression}

Recent advances in wireless foundation models have demonstrated substantial improvements in task performance and generalization. However, models with tens to hundreds of millions of parameters, and beyond, are often incompatible with the resource and latency constraints of wireless deployments. For instance, foundation models such as WirelessGPT~\cite{yang2025wirelessgpt} and EMind~\cite{luo2025emind} are pretrained at large scales, with initial parameter counts of 80 M and 110 M, respectively, while WiFo-M$^2$~\cite{zhang2026wifom2plugandplaymultimodalsensing} achieves multi-modal integration with under 10 M parameters.

Practical wireless applications typically require inference within microseconds to milliseconds, and available memory is measured in megabytes, not gigabytes. Accordingly, model compression is essential for bridging the gap between state-of-the-art architectures and real-world deployment. This section reviews prevalent compression methods, analyzes their trade-offs, and discusses their suitability for wireless physical-layer workloads. 

\subsection{Quantization}
\label{subsec:quantization}

Quantization reduces the numerical precision of model weights and activations,
yielding direct reductions in memory footprint, memory bandwidth, and
arithmetic cost.

\paragraph{Post-training quantization (PTQ)}
PTQ converts a trained FP32 model to lower precision (typically INT8) using
a small calibration dataset and no further gradient-based optimization.  Erak and Abou-Zeid~\cite{erak2023accelerating} demonstrate the effectiveness of PTQ for massive MIMO CSI feedback, where post-training dynamic range quantization combined with weight clustering achieves 86.5\% model size reduction and 76.2\% inference time reduction with minimal reconstruction quality loss. PTQ is therefore a low-cost initial compression step when calibration preserves the target-task accuracy.

\paragraph{Quantization-aware training (QAT)}
When PTQ is insufficient, particularly at INT4 or lower bit-widths needed
for FPGA and ASIC deployments, QAT embeds quantization operators into the
training graph so that the model learns to compensate for discretization
error.  Yellapragada et al.~\cite{yellapragada2025efficient} demonstrate that QAT-trained neural receivers at 4-bit and 8-bit precision achieve block error rates (BLER) comparable to full-precision FP32 models at 10\% target BLER, yielding 8$\times$ compression where PTQ alone fails.  For sub-millisecond tasks such as channel estimation and interference
detection, where models must run on fixed-point datapaths with 4--8-bit
arithmetic, QAT is essential.  Mixed-precision strategies, which assign
higher precision to sensitivity-critical layers (e.g., the output projection
of an attention block) and lower precision elsewhere, provide a fine-grained
accuracy--efficiency trade-off.  Beyond standard fixed-point QAT, Gupta et al.~\cite{gupta2025neuromorphicrx} demonstrate that coupling QAT with neuromorphic spiking architectures can yield 7.6$\times$ energy reduction for 5G-NR OFDM reception while maintaining competitive performance, illustrating the synergy between quantization and domain-specific compute paradigms.

\paragraph{Wireless-specific considerations}
Wireless signals are complex-valued (in-phase and quadrature components), and the dynamic range of channel coefficients can vary significantly across propagation environments.  Per-channel or per-token dynamic quantization scales, rather than a single global scale, are therefore important for
maintaining accuracy.  Block floating-point (BFP) formats, where groups of values share an exponent~\cite{drumond2018training}, offer a hardware-friendly compromise between per-element and per-tensor quantization and are particularly suited to wireless workloads where channel coefficients within a coherence block share similar magnitudes.

For \glspl{wpfm}, quantization is applied after pretraining or downstream adaptation, using calibration data that span the target bands, \gls{snr} levels, and channel scales. WiFo-2~\cite{liu2025foundationmodelintelligentwireless} validates this procedure: its 16.46\,M-parameter backbone is calibrated to INT8 with TensorRT and deployed on a Jetson AGX Orin with essentially unchanged channel-estimation and prediction performance.

\subsection{Pruning}
\label{subsec:pruning}

Pruning removes redundant parameters from a trained model to reduce both
storage and computation.

\paragraph{Structured pruning}
Structured pruning eliminates entire architectural units: attention heads, convolutional filters, or full Transformer layers.  The resulting model has a regular, dense computation graph that can be directly accelerated on ASICs and FPGAs without special sparse-compute hardware. For Transformer-based \glspl{wpfm}, attention-head pruning is particularly natural. Lu et al.~\cite{lu2025fcos} demonstrate structured pruning for automatic modulation recognition, combining channel-level and layer-level pruning to achieve 95.51\% FLOPs reduction and 95.31\% parameter reduction with only 0.46\% accuracy drop. Similarly, Erak and Abou-Zeid~\cite{erak2023accelerating} report that structured pruning of CSI feedback networks yields over 50\% compression with minimal quality degradation, confirming that wireless models contain substantial redundancy amenable to structured removal.

\paragraph{Unstructured pruning}
Unstructured (weight-level) pruning can achieve much higher sparsity
(80--95\%) but produces irregular memory access patterns that only benefit
performance on hardware with explicit sparse-tensor support (e.g., NVIDIA
Ampere's structured 2:4 sparsity, or dedicated sparse accelerators).  For
wireless applications running on general-purpose FPGAs, structured pruning
is almost always preferable.

\paragraph{Dynamic and input-dependent pruning}
Some wireless inputs are inherently sparse or bursty (e.g., spectrum sensing in lightly occupied bands).  Input-dependent pruning, where the model dynamically skips computation for uninformative regions, can yield significant savings in average-case FLOPs while preserving worst-case
accuracy.  This is closely related to token pruning, discussed below.

For \glspl{wpfm}, pruning can remove complete attention heads, feed-forward channels, experts, or layers after downstream adaptation. Importance should be measured across representative bands, antenna configurations, and environments before brief retraining, so that capacity used only under rare channel conditions is not removed.

\subsection{Knowledge Distillation}
\label{subsec:distillation}

Knowledge distillation trains a compact student model to replicate the
behavior of a large teacher model, transferring the teacher's learned
representations into a deployment-friendly form~\cite{gou2021knowledge}.

\paragraph{Task-specific distillation}
A practical deployment strategy for wireless foundation models is to pretrain a large, general-purpose teacher model using diverse wireless data (encompassing multiple frequency bands, environments, and hardware configurations), and subsequently distill task-specific student models~\cite{tang2019distilling}. Students are tailored for individual tasks such as channel estimation, positioning, and signal classification, and can be independently optimized for their respective hardware targets (e.g., INT4 on FPGA for channel estimation, INT8 on embedded GPU for signal classification).  Tavakolian et al.~\cite{tavakolian2026knowledge} demonstrate this paradigm for mmWave beam prediction: a compact student model retaining only two hidden layers of 64 neurons closely mimics a large sub-6\,GHz-to-mmWave teacher, reducing trainable parameters and FLOPs by over 99\% with only a 2.5 percentage point drop in top-1 beam selection accuracy compared to the teacher.

In the \glspl{wpfm} area, the student can match both task outputs and intermediate wireless representations from the shared backbone. Distillation can also reduce iterative inference: WiFo-MUD~\cite{yang2026wifomudwirelessfoundationmodel} distills a multi-step diffusion teacher into a single-step student while retaining comparable demodulation accuracy.

\paragraph{Feature-level vs.\ logit-level distillation}

Distillation can be performed at two levels. Logit-level distillation aligns the output distributions of the student and teacher models, focusing on the final predictions~\cite{jin2023multi}. Feature-level distillation, on the other hand, matches the activations of intermediate layers between teacher and student. These approaches differ in the granularity of knowledge transferred, and the choice between them may depend on the requirements of the target task~\cite{park2019feed}.  Ma et al.~\cite{ma2025knowledge} apply attention-based feature-level distillation to mmWave beam tracking, compressing a large attention-based teacher into a student that operates with 60\% shorter input sequences and over 16$\times$ fewer parameters while maintaining above 93\% tracking accuracy, demonstrating that feature-level transfer is particularly effective when the teacher encodes temporal patterns critical for beam coherence.

\paragraph{Heterogeneous distillation}
In distributed wireless systems (e.g., multi-agent cooperative sensing or
device--edge deployments), different nodes may run different hardware.
Heterogeneous distillation allows a single teacher to produce students
with different architectures (CNN for FPGA, Transformer for GPU), each
tailored to its target platform while sharing the same underlying knowledge.
While cross-architecture distillation techniques have been developed for vision tasks~\cite{yang2025multi}, their application to wireless foundation models, where device heterogeneity is the norm rather than the exception, remains an open opportunity.

\paragraph{Adversarial robustness transfer}
Standard distillation transfers accuracy but not necessarily robustness to
adversarial inputs.  For security-sensitive wireless applications, adversarial distillation, in which the teacher's robust decision boundaries are explicitly mimicked by the student under adversarial perturbations, is critical, as a distilled student may otherwise be substantially more vulnerable to adversarial RF attacks than its teacher~\cite{catak2022defensive}. We discuss adversarial robustness in greater depth in Section~\ref{subsec:adversarial}.

\subsection{Neural Architecture Search}
\label{subsec:nas}

Hardware-aware NAS automates the search for model topologies that meet
specific hardware constraints (latency, memory, energy) while maximizing
task accuracy.  In the WPFM context, NAS is most valuable at the
fine-tuning stage: given a pretrained backbone, NAS selects the minimal
sub-network (subset of layers, heads, and channels) that satisfies the
deployment constraints of a particular task and hardware target.  This
avoids the need for manual architecture design for each deployment scenario
and can discover non-obvious design points (e.g., a 3-layer, 4-head
Transformer with mixed INT4/INT8 precision) that outperform hand-tuned
baselines.  Auto-CsiNet~\cite{li2023autocisnet} applies gradient-descent-based NAS to massive MIMO CSI feedback, automatically generating scenario-customized architectures that achieve approximately 14\% improvement in reconstruction performance and 50\% complexity reduction compared to manually designed models, validating that NAS can effectively navigate the wireless-specific design space.

The pretrained backbone for \glspl{wpfm} can serve as a supernetwork from which NAS selects task-specific layers, heads, channels, or adapters under the memory and latency limits in Eq.~\eqref{eq:max_params}. The search should evaluate multiple wireless configurations to avoid selecting a subnetwork that is efficient but specific to one site or band.

\subsection{Token and Patch Pruning}
\label{subsec:token_pruning}

ViT and MAE-based \glspl{wpfm} that process spectrograms, CSI matrices, or radio
maps as 2D patches can benefit from dynamic token pruning.  A lightweight
scoring module evaluates each patch at an early layer and discards those
that carry little task-relevant information, for example, noise-only
spectral bins in spectrum sensing, or spatially featureless regions in
radio maps.  In vision Transformers, token pruning has been shown to reduce FLOPs by 30--60\% with minimal accuracy impact~\cite{rao2021dynamicvit}; wireless signals, which often contain spatially sparse information (e.g., few active users in a wideband spectrum), are expected to benefit similarly, though systematic evaluations on \glspl{wpfm} remain limited.

In a \gls{wpfm}, the scorer operates after tokenization and can combine learned task relevance with physical cues such as pilot locations and channel variation. CSI-JEPA~\cite{luo2026csijepafoundationrepresentationsubiquitous} demonstrates variation-based scoring of temporal-subcarrier patches for masking; adapting this idea to inference-time pruning remains an open direction.

\subsection{Early-Exit Architectures}
\label{subsec:early_exit}

Not all inputs require the full depth of a deep model. Early-exit architectures attach auxiliary classification or regression heads to intermediate layers, allowing the model to output a prediction and terminate computation when a confidence threshold is met. For wireless tasks with variable accuracy requirements, pedestrian localization (meter-level) vs.\ industrial AGV tracking (centimeter-level), or coarse modulation detection vs.\ fine-grained emitter fingerprinting, this provides a principled mechanism to trade latency for precision on a per-input basis.  Jankowski et al.~\cite{jankowski2023adaptive} develop early exiting for collaborative inference at wireless network edges, where a transmission-decision mechanism evaluates whether predictions from early exits should be finalized locally or forwarded to edge servers based on wireless channel conditions.  Zhang et al.~\cite{zhang2026channeladaptive} further propose channel-adaptive early exit, jointly adapting feature compression and model complexity to maximize edge processing throughput under latency and accuracy constraints.  These works demonstrate that early-exit mechanisms can substantially reduce communication overhead (by avoiding unnecessary transmissions for high-confidence local predictions) and improve edge processing throughput (up to 2$\times$ in favorable channel conditions~\cite{zhang2026channeladaptive}), while preserving full-model accuracy for difficult inputs. For \glspl{wpfm}, lightweight task heads can be attached to intermediate backbone layers during adaptation. An input exits when calibrated confidence satisfies the task requirement; low-\gls{snr} or out-of-distribution inputs continue through the full backbone, although this strategy has not yet been systematically evaluated for \glspl{wpfm}.

\subsection{Model Partitioning and Offloading}
\label{subsec:partitioning}

When no single device can host an entire foundation model, partitioning
the model across a device--edge or device--edge--cloud hierarchy distributes
the computational burden.  Wu et al.~\cite{wu2024device} propose a
device-edge cooperative fine-tuning framework in which adapters are
kept and updated on-device while data embeddings are transmitted
over the air via AirComp, and the heavy backbone resides
on the edge server. For inference, early layers can run on-device (extracting
local features from raw IQ or CSI data), and deeper layers execute on the
edge, with only compact intermediate embeddings transmitted over the wireless
link.  The partition point must balance three factors: (i)~on-device compute
and memory, (ii)~wireless link bandwidth and latency, and (iii)~privacy,
since early-layer activations may leak sensitive information about the
raw signal.  In multi-platform scenarios (e.g., UAV swarms or
distributed sensor networks), intermittent connectivity adds further
constraints on the partition design.

\subsection{Hardware--Software Co-Design}
\label{subsec:codesign}

The most aggressive efficiency gains come from jointly designing the model
architecture and the target hardware.  Unlike general-purpose AI
accelerators, wireless baseband processors can be customized for domain-specific
operations: custom number formats tuned to the dynamic range of channel
coefficients, fused attention kernels matching the antenna--subcarrier
structure of OFDM, and on-chip buffers sized to the model's activation
footprint.  While this approach requires significant engineering investment,
it is the most viable path for deploying foundation-model-derived
intelligence in the innermost loops of future 6G baseband processors,
and is particularly relevant to programs requiring custom silicon
for domain-specific wireless processing.

For \glspl{wpfm}, co-design jointly selects patch dimensions, numerical precision, on-chip activation buffers, and kernels for attention or \glspl{ssm}. Hardware intended for test-time adaptation must also support lightweight gradient computation; current forward-only edge accelerators make such adaptation more practical at GPU-equipped base stations~\cite{cheng2026bigwirelessfoundationmodel}.

\subsection{Recommended Compression Pipeline}
\label{subsec:compression_pipeline}

Based on the methods surveyed above, we outline a staged compression pipeline for practitioners deploying \glspl{wpfm}:

\begin{enumerate}
    \item \textbf{Stage~1 (Apply):} INT8 PTQ yields approximately 4$\times$ memory reduction (FP32$\to$INT8) and requires no retraining; Erak and Abou-Zeid~\cite{erak2023accelerating} confirm minimal quality degradation for CSI feedback at this precision.
    \item \textbf{Stage~2 (Latency-constrained):} Structured pruning (e.g., over 50\% compression as demonstrated by~\cite{erak2023accelerating}) or knowledge distillation (up to 99\% parameter reduction for task-specific students~\cite{tavakolian2026knowledge}) depending on whether the target task is known at compression time.
    \item \textbf{Stage~3 (Aggressive targets):} QAT to INT4 for FPGA/ASIC datapaths, combined with token pruning and early-exit for dynamic input-dependent savings.
    \item \textbf{Stage~4 (Custom hardware):} Hardware--software co-design with domain-specific number formats and fused operators for innermost baseband processing loops.
\end{enumerate}

The stages are cumulative: for example, Lyu et al.~\cite{lyu2026swiftchannel} combine QAT with knowledge distillation and FPGA co-design for 5G channel estimation, achieving 24$\times$ speed-up and 33$\times$ energy efficiency improvement over GPU baselines on Zynq hardware.

\section{Cross-Cutting Challenges and Future Research Directions}
\label{sec:challenges}
\label{sec:future}

The development of \glspl{wpfm} remains at an early stage. Current studies demonstrate transferable representations and multi-task adaptation, but most evaluations cover a limited set of datasets, configurations, and hardware platforms. The challenges below cut across telecommunications, localization, and sensing and are also coupled: broader data and model diversity can improve transfer while increasing training cost, deployment complexity, and the security and privacy risks of a shared backbone. We therefore pair each limitation below with the research directions needed to address it.

\subsection{Data, Benchmarks, and Scaling Laws}
\label{subsec:data}
\label{subsec:future_scaling}

Wireless datasets differ in frequency, bandwidth, antenna geometry, mobility, propagation environment, hardware, and task labels. Synthetic data provide coverage at low cost but do not fully reproduce hardware impairments and environmental dynamics, while measured datasets are expensive and often narrow in scope. CeBed and DeepTelecom provide useful components for channel estimation and multimodal channel learning~\cite{feriani2023cebed,wang2025deeptelecom}, but a WPFM benchmark must connect task families through common metadata and evaluation rules. It should define held-out splits for sites, bands, arrays, hardware, mobility, and tasks; state the target-domain adaptation budget; and report data provenance, latency, energy, and failure cases.

These requirements are also necessary for studying scaling. Existing results vary across antenna and frequency extrapolation~\cite{jiang2026csimaemaskedautoencoderbasedchannel}, data composition~\cite{zhang2026hetercsichanneladaptiveheterogeneouscsi}, and channel intrinsic dimensionality~\cite{cheng2026bigwirelessfoundationmodel}. Future compute-optimal studies should jointly vary model size, corpus size, and training compute across multiple modalities, tasks, transfer axes, and architectures. They should report saturation and uncertainty and include inference-time adaptation in the resource budget. Until such evidence is available, parameter count alone is not a reliable design rule for WPFMs.

\subsection{Cross-Domain Generalization and Real-Time Adaptation}
\label{subsec:generalization}

Generalization must be evaluated across bands, antenna configurations, environments, mobility patterns, hardware, and tasks. Success on one axis does not imply robustness to compound shifts, such as a new site observed through a different array and frequency band. Multimodal context can reduce ambiguity~\cite{10971878}, but it creates dependence on sensor availability and alignment. Evaluations should distinguish strict zero-shot transfer from adaptation using target pilots, unlabeled samples, or labels because these protocols incur different costs.

Adaptation should also match the timescale of the underlying change. Pilot-aided \ac{ttt} can address slower shifts in geometry or environment, but current backpropagation-based methods are generally too slow for symbol-level or fast-fading adaptation~\cite{cheng2026bigwirelessfoundationmodel}. Promising directions include drift detection, adapter-only updates, retrieval from environment-specific memory, state-space updates, and periodic server-side adaptation followed by validated model delivery. Reports should include adaptation latency, pilot overhead, forgetting, and behavior when adaptation data are noisy or adversarial.

\subsection{Computational Cost, Hybrid Processing, and Real-World Deployment}
\label{subsec:compute}
\label{subsec:future_hybrid}

Pretraining requires sustained accelerator capacity, whereas real-time inference is constrained by on-chip memory, bandwidth, latency, and energy. This gap is especially large at the \gls{ue}, where a backbone with tens of millions of parameters cannot be placed directly in an inner baseband loop. WiFo-2 demonstrates INT8 deployment on an edge platform~\cite{liu2025foundationmodelintelligentwireless}, while lightweight and device--edge methods provide complementary paths for constrained devices~\cite{cheraghinia2025lightweightfoundationmodelwireless,10558820}. Nevertheless, feasibility must be established through end-to-end latency, energy, memory, buffering, and accuracy measurements rather than parameter count alone.

Replacing every block with a learned model is neither necessary nor always desirable. Hybrid systems can use model-based transforms before a WPFM, embed physical losses during pretraining, refine conventional estimates, or switch between learned and classical paths according to uncertainty and channel conditions. SPA-MAE demonstrates propagation-aware representation learning~\cite{chen2026spamaephysicsguidedcsifoundation}, while pilot-aided adaptation results show a regime-dependent crossover between least-squares interpolation and an adapted model~\cite{cheng2026bigwirelessfoundationmodel}. Future prototypes should identify stable interfaces and fallback criteria and compare hybrid, purely learned, and purely model-based systems under the same pilot and latency budgets. They must also address the full lifecycle of calibration, versioning, monitoring, rollback, and adaptation.

\subsection{Standardization and Interoperable WPFM Interfaces}
\label{subsec:future_standardization}

3GPP studies on AI-enabled air-interface functions emphasize procedures, interfaces, model lifecycle management, performance monitoring, and fallback rather than prescribing one neural architecture~\cite{lin2023overview3gppstudyartificial,9970357}. WPFMs add interoperability questions because a backbone may be shared across vendors, devices, and tasks. Candidate standardized boundaries include input tensors and tokenizer metadata, the dimensions and semantics of exchanged latent representations, task-head interfaces, model identifiers and versions, capability negotiation, and fallback procedures. Standardizing a complete tokenizer or latent space could support encoder--decoder interoperability for two-sided functions such as \gls{csi} feedback, but could also restrict innovation. A practical first step is to standardize observable behavior, metadata, and compatibility tests while allowing internal architectures to evolve.

\subsection{Trustworthiness, Security, and Privacy}
\label{subsec:adversarial}
\label{subsec:privacy}

A shared backbone creates a common failure point across downstream tasks. Attacks may target the signal, pretraining corpus, adaptation data, or model updates. Task-specific studies show that adversarial perturbations can degrade \gls{rf} fingerprinting and channel estimation~\cite{ma2025adversarial,catak2022defensive}, but it remains unclear whether attacks and defenses transfer across WPFM heads. Evaluation requires physically realizable threat models, calibrated uncertainty, attack transfer across tasks, recovery behavior, and a conventional fallback path.

Trustworthy models should also expose the signal regions or physical factors supporting a decision and check outputs against link budgets, spectral masks, protocol states, and feasible action sets. Neuro-symbolic methods can combine continuous \gls{rf} representations with explicit rules and regulatory constraints~\cite{fontaine2025reasoning}; the main open problem is a low-latency interface between distributed embeddings and discrete symbols. Privacy must be addressed across the same lifecycle. Federated pretraining retains raw data locally but faces heterogeneous data, intermittent participation, and poisoning~\cite{10558823}. Over-the-air aggregation can provide differential-privacy guarantees under specific assumptions~\cite{liang2025differential}, but secure aggregation, contribution auditing, leakage from embeddings, and deletion or unlearning remain open.

\subsection{Distributed, Cross-Layer, and Agentic Intelligence}
\label{subsec:distributed}
\label{subsec:llm_integration}

Partitioning a WPFM across users, base stations, edge servers, and the cloud can reduce local computation but introduces communication delay, synchronization errors, and failures under intermittent links. Existing distributed frameworks illustrate knowledge sharing, expert placement, and adaptive computation~\cite{11303197,10.1109/MWC.009.2300501,xue2024wdmoe,verbruggen2024computational}. A complete design must jointly optimize the partition point, transmitted representation, radio resources, tail latency, and fallback behavior.

Cross-layer and agentic systems extend this problem. Physical-layer representations could inform scheduling, routing, handover, and resource allocation, but these functions operate at different timescales and expose different objectives. A practical architecture may combine a fast WPFM encoder, layer-specific state abstractions, and a slower supervisory planner rather than one specific model. \Glspl{llm} can interpret intents and plan workflows, while WPFMs and conventional algorithms provide fast perception and control. WirelessAgent++, ComAgent, and related work illustrate workflow generation and multi-agent coordination~\cite{tong2025wirelessagent,li2025comagent,liang2025largelanguagemodelswireless}. Actions must be grounded in measured state, restricted to valid commands, and checked before execution; compact non-linguistic interfaces may reduce latency.

\subsection{Environment-Aware World Models and Non-Terrestrial Networks}

World models aim to learn how geometry, mobility, and electromagnetic propagation jointly determine future channel states. The WWM framework combines \gls{csi}, 3D point clouds, and trajectories to learn an environment-grounded predictive representation~\cite{chen2026wirelessworldmodelainative}, connecting WPFMs with digital twins and environment-intelligent communication~\cite{10148936,10742564,ZHANG2026186}. Open questions include how to infer environmental state from sparse radio observations, update it when scenes change, and test physical consistency rather than predictive accuracy alone.

Non-terrestrial networks provide a demanding test case because they introduce large Doppler shifts, long delays, moving coverage regions, and limited site-specific data. Initial results on simulated satellite channels suggest benefits from right-sized adaptive models~\cite{cheng2026bigwirelessfoundationmodel}, but evidence remains limited across real measurements, orbital regimes, and terrestrial--non-terrestrial transitions. Future datasets should combine orbital state, beam configuration, weather, and channel measurements and should distinguish geometry-level adaptation from fast-fading tracking.

\subsection{Research Priorities and Evaluation}

Near-term progress depends on comparable benchmarks, reproducible hardware measurements, explicit transfer protocols, and interoperable lifecycle procedures. World models, neuro-symbolic reasoning, cross-layer pretraining, and agentic control are longer-term directions whose value should be compared with simpler modular systems. Across all topics, the decisive question is not whether a WPFM can perform another task, but whether sharing a backbone improves transfer, lifecycle cost, and reliability under a fixed resource budget.


\section{Conclusion}
\label{sec:conclusion}

This paper reviews wireless physical-layer foundation models as a shift from isolated task-specific networks toward reusable representations learned from broad wireless data. It connects four design stages that are often treated separately: signal tokenization and neural architecture, self-supervised pretraining, downstream adaptation, and deployment under wireless hardware constraints. The proposed taxonomy organizes existing WPFMs by architecture, input modality, pretraining objective, model scale and deployment target, and generalization capability. Applications in telecommunications, localization, and sensing show that the same \gls{csi}, \gls{iq}, or spectrogram representation can support multiple tasks, but each domain emphasizes a different operational constraint: latency and signaling overhead in communication, cross-environment transfer in localization, and label scarcity in sensing.

The reviewed evidence suggests that wireless foundation models need not follow the parameter-scaling path of language models. Physics-aware tokenization, masking, multimodal context, sparse expert routing, state-space architectures, and test-time adaptation can be more important than model size alone. At the same time, reported gains are difficult to compare because datasets, configurations, adaptation budgets, and hardware measurements remain inconsistent. Model compression, distillation, partitioning, and hardware--software co-design can reduce deployment cost, but they must be evaluated together with generalization and robustness rather than as independent post-processing steps.

WPFMs are therefore best viewed as a developing systems technology rather than a finished replacement for conventional physical-layer algorithms. Progress toward practical use requires open cross-domain benchmarks, real-world validation, trustworthy and privacy-aware adaptation, and clear interfaces between shared backbones, specialized heads, conventional signal processing, and higher-level agents. If these requirements are addressed, WPFMs can provide a common representation layer for communication, localization, and sensing while reducing repeated data collection and model maintenance across heterogeneous wireless deployments.

\bibliographystyle{unsrt}
\bibliography{References}

\end{document}